\documentclass[prb,twocolumn,floats]{revtex4-2}
\usepackage{amsmath}
\usepackage{color}
\usepackage{graphicx}
\graphicspath{Figures}

\newcommand{\HH}{\mathbf{H}}

\newcommand{\br}{\mathbf{r}}
\newcommand{\bk}{\mathbf{k}}
\begin{document}
	
\title{Spin-orbit coupling and the Edelstein effect at conducting ferroelectric domain walls}
\author{Maryam A. Nasir and W. A. Atkinson$^\ast$}
\date{\today}
\affiliation{Department of Physics \& Astronomy, Trent University, Peterborough, Ontario K9L 0G2, Canada}
\email{billatkinson@trentu.ca}
\begin{abstract}
Head-to-head ferroelectric domain walls are intrinsically charged, and are typically compensated by a mix of oppositely charged defects and free electrons.  The free electrons  form a two-dimensional electron gas (2DEG) along the domain wall. In many cases, inversion symmetry is broken at the  wall, which implies that the 2DEG is subject to nontrivial spin-orbit coupling.  Here, we use symmetry arguments to construct a generic six-band tight-binding electronic Hamiltonian for a $90^\circ$ head-to-head ferroelectric domain wall.  The model, which includes spin-orbit physics and has a multi-orbital $t_{2g}$ band structure that is common to transition-metal perovskites, is applied to BaTiO$_3$.  We find that the 2DEG develops an Ising spin texture, with spins aligned perpendicular to the domain wall.  We contrast this with the Rashba spin texture that should emerge at weakly conducting $90^\circ$ head-to-tail domain walls. We then show that the head-to-head domain walls should have a measurable Edelstein effect (that is, a current-induced magnetization), even in the dilute limit and at room temperature, and describe a simple experiment to measure it.
\end{abstract}

\maketitle

\section{Introduction}

Although first predicted nearly 60 years ago \cite{Guro:1968,Guro:1970,Krapivin:1970,Vul:1973}, conducting domain walls (CDWs) in otherwise-insulating ferroelectrics were not observed experimentally until 2009 \cite{seidel2009conduction}.  Since then, domain-wall conductivity has been reproducibly observed in a variety of ferroelectric materials, including BiFeO$_3$ \cite{seidel2010domain,maksymovych2011dynamic,farokhipoor2011conduction,rojac2017domain,zhang2019intrinsic}, BaTiO$_3$ \cite{sluka2013free,Beccard:2022}, and LiNbO$_3$ \cite{schroder2012conducting,werner2017large,volk2017domain,Qian:2022,McClusky:2025}. 
The conductivity is generally largest for charged domain walls because the polarization discontinuity at the wall generates a bound charge that is compensated by some combination of impurities, defects, and holes or electrons (depending on the sign of the bound charge).  The electrons or holes may form a two-dimensional electron gas (2DEG) that is confined to the domain wall, which therefore acts as a conducting channel. The above reasoning suggests that neutral domain walls should be insulating; however, nominally neutral walls are found to be weakly conducting in, for example, BiFeO$_3$ \cite{zhang2019intrinsic}.

CDWs can be manipulated by external fields and strains, and 
for this reason have been explored as functional elements in reconfigurable circuits or memristors \cite{Sharma:2017,Ma:2018,Jiang:2018,Sharma:2019,McConville:2020,Risch:2022,Wang:2022,Liu:2023, Ratzenberger:2024}. However, progress towards device development is inhibited by an incomplete understanding of the fundamental physics of the domain walls themselves \cite{sharma2022roadmap}.  There remain, for example, foundational questions around domain formation \cite{sturman2023effect,Kosobokov:2024conductivity,bednyakov2025fragmented,yang2024intrinsic,marton2025pyramidal}, dynamics \cite{Khachaturyan:2024,carroll2025dynamics,Gurung:2025extrinsic,McCluskey:2025Review},  electronic structure \cite{sifuna:2020DFT,verhoff2024DFT}, and conduction mechanisms \cite{Lubk:2009DFT,zhang2019intrinsic,Risch:2022,Zahn:2024,yang2024intrinsic}.  In broad terms, these studies all address the question of how one can realiably create and control CDWs.

\begin{figure}[tb]
    \includegraphics[width=\columnwidth]{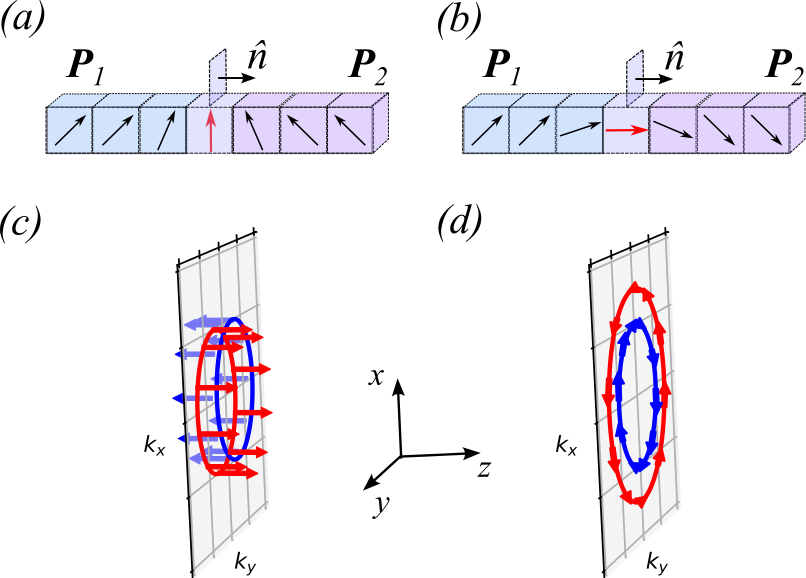}
    \caption{Polarization distributions and corresponding  Fermi surfaces for N\'eel domain walls.  Polarization vectors are shown for (a) charged head-to-head and (b) nominally neutral head-to-tail walls. A 2DEG is bound to each domain wall.  Their spin-resolved Fermi surfaces, shown in (c) and (d) respectively, are split by SOC, which by symmetry are of (c) Ising and (d) Rashba types.   Red and blue contours show the spin-split Fermi surfaces, and arrows indicate the direction of spin polarization at different points on the Fermi surfaces.}
    \label{fig:DWcartoon}
\end{figure}

Here, we ask a rather different question:  assuming that one can engineer an ideal two-dimensional conducting domain wall, how does the polarization profile affect the intrinsic properties of the 2DEG?  In particular, we focus on the emergence of nontrivial spin-orbit coupling at CDWs.
Figure~\ref{fig:DWcartoon} shows two domain-wall configurations that are relevant to our discussion.  Figure~\ref{fig:DWcartoon}(a) shows a head-to-head domain wall.  We assume that the wall is of the N\'eel type, namely that the polarization ${\bf P}$ rotates from ${\bf P}_1$ on the left to ${\bf P}_2$ on the right without leaving the plane defined by ${\bf P}_1$ and ${\bf P}_2$. This wall is charged, since $({\bf P}_1-{\bf P}_2)\cdot {\bf \hat n} \neq 0$ (${\bf \hat n}$ is the unit vector normal to the wall). In BaTiO$_3$, $90^\circ$ walls of this type are conducting \cite{sluka2013free}.  Figure~\ref{fig:DWcartoon}(b) shows a head-to-tail N\'eel-type wall; this is nominally neutral [$({\bf P}_1-{\bf P}_2)\cdot {\bf \hat n} = 0$], but similar $71^\circ$ walls are conducting in BiFeO$_3$ \cite{zhang2019intrinsic}. There, it is argued that the conduction mechanism is intrinsic rather than caused by defects or oxygen vacancies \cite{zhang2019intrinsic,Lubk:2009DFT}.  

In both cases, we have assumed that the domain wall has N\'eel character; however, the physics of Bloch-type domain walls (for which the polarization winds out of the plane of ${\bf P}_1$ and ${\bf P}_2$) will be similar. The key point is that the polarization profile in the wall breaks the inversion symmetry of the 2DEG, which necessarily introduces spin-orbit coupling (SOC). Conversely, walls along which the polarization vanishes should have only weak SOC. We show below that, for the head-to-head wall  in Fig.~\ref{fig:DWcartoon}(a), symmetry arguments constrain the SOC to be of Ising type, with the spin axis perpendicular to the 2DEG; for the head-to-tail wall shown in Fig.~\ref{fig:DWcartoon}(b), the SOC is of the Rashba type, with spins in the plane of the wall.  The 2DEG conduction bands therefore have distinct $k$-space spin textures that depend on the polarization configuration at the wall; these are illustrated in Figs.~\ref{fig:DWcartoon}(c) and (d), respectively, for the simplest possible two-band models (with one orbital and two spin degrees of freedom).  Because a nearly-identical Rashba case has been discussed at length in the context of oxide interfaces \cite{zhong2013theory,tao2016strain,bruno2019band,johansson2021spin,kumar2022rashba,trama2022tunable}, we focus primarily on the effects of Ising SOC at head-to-head walls.

Although illustrative, two-band models are inadequate for  transition-metal perovskites.  In these materials, the electronic conduction bands are derived from the $t_{2g}$ orbitals of the B cation (Ti or Fe in BaTiO$_3$ and BiFeO$_3$, respectively).  Spin-orbit terms are off-diagonal in this basis, and the conduction bands contain nontrivial admixtures of the $d_{xy}$, $d_{yz}$, and $d_{zx}$ orbitals.  Consequently, the electronic conduction bands have both spin and orbital angular momentum textures in $k$-space.

These angular momentum textures have observable consequences and, we propose, raise the possibility of using CDWs as reconfigurable circuit elements in so-called ``spin-orbitronics'' applications.  The simplest applications involve the interconversion between charge currents and magnetization via the Edelstein or inverse Edelstein  effects \cite{han2018quantum}.  Importantly, large spin-to-charge conversion efficiencies have been observed at KTaO$_3$ \cite{vicente2021spin,al2025spin,zhang2019thermal} and especially SrTiO$_3$ \cite{lesne2016highly,song2017observation,noel2020non,el2023observation} interfaces.   Here we show that similar efficiencies may be attainable for CDWs in transition-metal perovskite ferroelectrics.

In Sec.~\ref{sec:calculations}, we obtain model tight-binding Hamiltonians for the 2DEGs pictured in Figs.~\ref{fig:DWcartoon}.  Our approach is to use symmetry arguments to constrain the form of the Hamiltonian matrix.  This leaves a large number of unknown parameters; however, only a few of them are relevant to spin-orbit physics.  Their influence on the $k$-space spin textures is discussed in Sec.~\ref{sec:results}.  To connect our models to other published work, we then  make  semi-classical estimates of the Edelstein conversion efficiencies, which determine the magnitude of the magnetization induced by an applied electric field.  We find that, for parameters appropriate to BaTiO$_3$, the field-induced current should be observably spin polarized.

\section{Calculations}
\label{sec:calculations}
\subsection{Formal Considerations}
In this section, we discuss the symmetry constraints for an electron gas confined to the domain walls pictured in Figs.~\ref{fig:DWcartoon}(a) and (b).  For simplicity, we consider the limit in which the domain wall is only a single unit cell wide; in this case, the 2DEG comprises six bands derived from three orbital ($d_{xy}$, $d_{yz}$, and $d_{zx}$) and two spin flavours. In thicker domain walls, there will be multiple sub-bands of each flavour \cite{Gureev:2011,sturman_quantum_2015,Chapman:2022,atkinson2022evolution,cornell2023influence}, and our model would then apply to low electron densities where only the lowest-energy sub-band of each type is occupied.  We stress that the principal goal of this work is to demonstrate the possibility of significant spin-orbit physics at CDWs, and to motivate more sophisticated future calculations. As a further simplification, we limit the discussion to domain walls with normals along the cubic $\langle 001 \rangle$ directions, for which the $t_{2g}$ band structure is straightforward to obtain.  This latter assumption affects the orbital character of the Fermi surfaces, but does not change the essential spin-orbit physics.

To describe the 2DEG, we define an orthonormal basis of two-dimensional Bloch states 
\[
|\alpha \bk\rangle = \frac{1}{\sqrt{N}}\sum_i e^{i\bk \cdot \br_i} \phi_\alpha(\br-\br_i)
\]
where $\alpha \in \{ xy,\, yz,\, zx \}$ are the orbital labels,  $\phi_\alpha(\br-\br_i)$ is a localized orbital centered at lattice site $i$, and there are $N$ lattice sites.  The position $\br$ is a three dimensional vector, but the lattice sites $\br_i$ and wavevector $\bk$ lie in the xy plane.  

In this basis, the Hamiltonian can be represented by a $6\times 6$ matrix, with three orbital and two spin degrees of freedom,
\begin{equation}
[\HH_\bk]_{\alpha \chi, \beta \chi'} = \langle \alpha \bk|\langle \chi| \hat H |\chi'\rangle |\beta \bk \rangle
\label{eq:Hk1}
\end{equation}
where $|\chi \rangle$ is a two-component spinor and $\hat H$ is the operator form of the Hamiltonian.  It is useful to write the  matrix as 
\begin{equation} 
    \HH_\bk = \mathbf{H}_{0,\bk} \otimes \sigma_0 + \mathbf{H}_{1,\bk} \otimes \sigma_1 + \mathbf{H}_{2,\bk} \otimes \sigma_2 + \mathbf{H}_{3,\bk} \otimes \sigma_3
    \label{eq:Hexpand}
\end{equation}
where $\sigma_\mu$, $\mu = 0,\ldots,3$, are the Pauli matrices representing spin operators, and $\mathbf{H}_{\mu,\bk}$ are $3\times 3$ matrices representing the orbital degrees of freedom.  The orbital matrices satisfy $\HH_{\mu,\bk} = \frac 12 \mathrm{Tr}_\sigma \, [\HH_\bk \sigma_\mu]$, with $\mathrm{Tr}_\sigma$ the trace over spin degrees of freedom.

If the Hamiltonian is invariant under a unitary symmetry operation $\hat U = \hat U_\mathrm{orb}\hat U_\mathrm{sp}$, with $\hat U_\mathrm{orb}$ and $\hat U_\mathrm{sp}$ acting on orbital and spin degrees of freedom, respectively, then $[\HH_\bk]_{\alpha\chi,\beta\chi'}$ must be equal to the Hamiltonian matrix element $[\tilde \HH_\bk]_{\alpha\chi,\beta\chi'}$ constructed from transformed basis states,
\begin{eqnarray}
[\tilde \HH_\bk]_{\alpha \chi, \beta \chi'} &=& \langle  \hat U_\mathrm{orb} \alpha \bk|\langle \hat U_\mathrm{sp} \chi|  \hat H |\hat U_\mathrm{sp} \chi'\rangle | \hat U_\mathrm{orb} \beta \bk \rangle
\nonumber \\
&=& \sum_{\mu=0}^3 [\tilde \HH_{\mu,\bk}]_{\alpha\beta} \otimes \sigma_{\mu,\chi\chi'},
\label{eq:Hk2}
\end{eqnarray}
with
\begin{equation}
[\tilde \HH_{\mu,\bk}]_{\alpha\beta}
=  \frac 12 \langle \hat U_\mathrm{orb} \alpha \bk| \mathrm{Tr}_\sigma\left[
\hat H\cdot(\hat U_\mathrm{sp} \sigma_\mu \hat U_\mathrm{sp}^\dagger) \right ] | \hat U_\mathrm{orb}\beta \bk \rangle. 
\end{equation}
For the domain walls shown in Fig.~\ref{fig:DWcartoon}, the relevant unitary symmetries are reflections and  $C_4$ rotations about the z axis (in the Rashba case). For the reflections,  
$\hat U_\mathrm{sp}\sigma_\mu \hat U_\mathrm{sp}^\dagger = \pm \sigma_\mu$, where the sign depends on both the reflection plane and $\mu$.
Equivalence of  $\HH_\bk$ and $\tilde \HH_\bk$ under reflections thus requires that 
\begin{equation}
\tilde \HH_{\mu,\bk} = \pm \HH_{\mu,\bk},
\label{eq:Hmu1}
\end{equation}
For $C_4$ rotations $\sigma_\mu \stackrel{C_4}{\rightarrow} \pm \sigma_\nu$, and equivalence under rotations requires
\begin{equation}
    \tilde \HH_{\nu,\bk} = \pm \HH_{\mu,\bk}.
\end{equation}
For rotations around the z-axis, 
$(\sigma_0, \sigma_1, \sigma_2, \sigma_3) \stackrel{C_4}{\rightarrow} (\sigma_0, \sigma_2, -\sigma_1, \sigma_3)$.
A similar relation can be obtained for the anti-unitary time-reversal (TR) symmetry,
\begin{equation}
    [\HH_{\mu,\bk}]_{\alpha\beta} = \pm [\HH_{\mu,-\bk}]_{\alpha\beta}^\ast
\end{equation}
where the $+$ sign holds for $\mu=0$ and the $-$ sign for $\mu=1,\, 2,\, 3$.

\subsection{Two-band model for a head-to-head domain wall}
\label{sec:twoband}
As a simple illustration, we consider the head-to-head domain wall pictured in Fig.~\ref{fig:DWcartoon}(a) for the case of a single isotropic orbital per unit cell.  
We assume the 2DEG lies in the xy plane, and that it has the bare band structure $\epsilon_{0,\bk} = \hbar^2(k_x^2+k_y^2)/2m$.  Here, Eq.~(\ref{eq:Hexpand}) reduces to
\begin{equation}
H^I_\bk = \epsilon_{0,\bk} \sigma_0 + \epsilon_{1,\bk} \sigma_1 + \epsilon_{2,\bk} \sigma_2+ \epsilon_{3,\bk} \sigma_3,
\end{equation}
where the superscript $I$ indicates that the SOC is of the Ising type, and where $\epsilon_{\mu,\bk}$ are simple functions of $\bk$. At the domain wall, the ferroelectric polarization vector lies in-plane, with ${\bf P} = P_x {\bf \hat x}$. In addition to TR symmetry, the system has reflection symmetries $\hat M_z$ and $\hat M_y$ in the xy  and xz planes, respectively,  and reflection symmetry $\hat M_x$ in the yz plane. The latter is understood to transform the $x$-component of the polarization $P_x\rightarrow -P_x$ as well as the electronic states.  We can use these symmetries to obtain a generic Hamiltonian,
\begin{itemize}
    \item Under $\hat M_z$, $\sigma_{1,2} \rightarrow -\sigma_{1,2}$ and $\sigma_{0,3} \rightarrow \sigma_{0,3}$.  Invariance of $H^I_\bk$ requires $\epsilon_{1,\bk}=\epsilon_{2,\bk}=0$.
    \item Under $\hat M_y$, $\sigma_3\rightarrow-\sigma_3$ and $\epsilon_{3,(k_x,k_y)}\rightarrow \epsilon_{3,(k_x,-k_y)}$.  Invariance of $H^I_\bk$ requires that $\epsilon_{3,\bk}$ is odd in $k_y$.
    \item Under $\hat M_x$, $P_x\rightarrow-P_x$, $\sigma_3\rightarrow -\sigma_3$, and $k_x\rightarrow -k_x$.  Invariance of $H^I_\bk$ requirs that $\epsilon^3_{\bk}$ be a sum of terms that are odd in either $k_x$ or $P_x$.
    \item Under TR, $\sigma_3 \rightarrow-\sigma_3$ and $\epsilon_{3,\bk} \rightarrow \epsilon_{3,-\bk}^\ast$.  Since Hermiticity requires  that $\epsilon_{3,\bk}$ be real,  invariance of $H^I_\bk$ requires $\epsilon_{3,\bk}$ be odd in $\bk$.
\end{itemize}
To linear order in $\bk$ and $P_x$, the most general form for $\epsilon_{3,\bk}$ is thus
$\epsilon_{3,\bk} = \alpha P_x k_y$
and 
\begin{equation}
H^I_\bk = \frac{\hbar^2k^2}{2m} \sigma_0 + \alpha_I P_x k_y \sigma_3.
\label{eq:HI1bd}
\end{equation}
The Fermi surfaces and spin textures generated by this Hamiltonian are pictured in Fig.~\ref{fig:DWcartoon}(c).

A similar analysis for the head-to-tail domain wall leads to the Hamiltonian,
\begin{equation}
    H^R_\bk = \frac{\hbar^2k^2}{2m} \sigma_0 + \alpha_R P_z k_y \sigma_1 - \alpha'_R P_z k_x \sigma_2.
\end{equation}
In the limit that the 2DEG is a single unit cell wide, there is an addition $C_4$ rotational symmetry that enforces $\alpha_R = \alpha'_R$.  The Fermi surfaces and spin textures generated by this Hamiltonian are pictured in Fig.~\ref{fig:DWcartoon}(d).

\subsection{Six-band model for a head-to-head domain wall}
We now consider the head-to-head domain wall shown in Fig.~\ref{fig:DWcartoon}(a) for the case in which the conduction bands are formed from a single set of $t_{2g}$ orbitals per unit cell. The analysis of the previous section must be expanded to include the orbital symmetries.  For example, under  $\hat M_z$, the Bloch states transform as $\hat M_z |xy \bk\rangle \rightarrow |xy \bk\rangle$, $\hat M_z |yz \bk\rangle \rightarrow -|yz \bk\rangle$, and $\hat M_z |zx \bk\rangle \rightarrow - |zx \bk\rangle$.  Equation~(\ref{eq:Hmu1}) then becomes
\begin{eqnarray}
[\HH^I_{\mu,\bk}]_{xy,yz} &=& \frac 12 \left \langle xy\bk \left | \mathrm{Tr}_\sigma\left [ \hat H^I \sigma_\mu \right ] \right |yz \bk \right \rangle \nonumber \\
&=& \frac 12\left \langle \hat M_z xy\bk \left | \mathrm{Tr}_\sigma\left [ \hat H^I \hat M_z \sigma_\mu \hat M_z^{-1} \right ] \right |\hat M_z yz \bk \right \rangle \nonumber \\
&=& \mp\frac 12 \left \langle xy\bk \left | \mathrm{Tr}_\sigma\left [ \hat H^I \sigma_\mu \right ] \right |yz \bk \right \rangle,
\end{eqnarray}
where the $-$ ($+$) sign holds for $\mu = 0,\, 3$ ($\mu=1,\, 2$).  From this, we obtain $[\HH_{0,\bk}]_{xy,yz} = [\HH_{3,\bk}]_{xy,yz} = 0$.

Repeating this process for all symmetries, we obtain
\begin{eqnarray}
    \HH^I_{0,\bk} &=& \left[ \begin{array}{ccc}
    \epsilon^{xy}_\bk & 0 & 0 \\
    0 & \epsilon^{yz}_\bk & t^{yz,zx}_{0,\bk} \\
    0 & (t^{yz,zx}_{0,\bk})^\ast & \epsilon^{zx}_\bk
    \end{array} \right ] \\
    \HH^I_{1,\bk} &=& \left[ \begin{array}{ccc}
    0 & t_{1,\bk}^{xy,yz} & -i\frac{\xi}2 + t_{1,\bk}^{xy,zx} \\
    (t_{1,\bk}^{xy,yz})^\ast & 0 & 0 \\
    i\frac{\xi}2+ (t_{1,\bk}^{xy,zx})^\ast & 0 & 0
    \end{array} \right ] \\
    \HH^I_{2,\bk} &=& \left[ \begin{array}{ccc}
    0 & i\frac{\xi}2 + t_{2,\bk}^{xy,yz} & t_{2,\bk}^{xy,zx} \\
    -i\frac{\xi}2 + (t_{2,\bk}^{xy,yz})^\ast & 0 & 0 \\
    (t_{2,\bk}^{xy,zx})^\ast & 0 & 0
    \end{array} \right ] \\
    \HH^I_{3,\bk} &=& \left[ \begin{array}{ccc}
    0& 0 & 0 \\
    0 & 0 & i\frac{\xi}{2}+t^{yz,zx}_{3,\bk} \\
    0 & -i\frac{\xi}2+(t^{yz,zx}_{3,\bk})^\ast & 0
    \end{array} \right ] 
\end{eqnarray}
where $\xi$ is the atomic spin-orbit coupling, and the remaining nonzero terms are constrained to satisfy the symmetry relations shown in Table~\ref{Table:1}.  

\begin{table}[]
    \begin{tabular}{c|c|c|c}
    Element & $\hat M_x$  & $\hat M_y$ & TR \\
    \hline
        $\epsilon^\alpha_\bk$ & $\epsilon^\alpha_{(-k_x,k_y)}(-P_x)$ & $\epsilon^\alpha_{(k_x,-k_y)}$ &  $\epsilon^\alpha_{-\bk}$ \\
        $t_{0,\bk }^{yz,zx}$ &  $-t_{0,(-k_x,k_y) }^{yz,zx}(-P_x) $ & $-t_{0,(k_x,-k_y) }^{yz,zx} $ & $\left [ t_{0, -\bk}^{yz,zx} \right ]^\ast $ \\
        $t_{1,\bk }^{xy,yz}$ &  $-t_{1,(-k_x,k_y) }^{xy,yz} (-P_x)$ & $-t_{1,(k_x,-k_y) }^{xy,yz} $ & $ \left [ -t_{1, -\bk}^{xy,yz} \right ]^\ast $ \\
        $t_{1,\bk }^{xy,zx}$ &  $t_{1,(-k_x,k_y) }^{xy,zx} (-P_x)$ & $t_{1,(k_x,-k_y) }^{xy,zx} $ & $ \left [ -t_{1, -\bk}^{xy,zx} \right ]^\ast $ \\
        $t_{2,\bk }^{xy,yz}$ &  $t_{2,(-k_x,k_y) }^{xy,yz} (-P_x)$ & $t_{2,(k_x,-k_y) }^{xy,yz} $ & $ \left [ -t_{2, -\bk}^{xy,yz} \right ]^\ast $ \\
        $t_{2,\bk }^{xy,zx}$ &  $-t_{2,(-k_x,k_y) }^{xy,zx} (-P_x)$ & $-t_{2,(k_x,-k_y) }^{xy,zx} $ & $ \left [ -t_{2, -\bk}^{xy,zx} \right ]^\ast $ \\
        $t_{3,\bk }^{yz,zx}$ &  $t_{3,(-k_x,k_y) }^{yz,zx} (-P_x)$ & $t_{3,(k_x,-k_y) }^{yz,zx} $ & $\left [ -t_{3, -\bk}^{yz,zx} \right ]^\ast $ \\
        \hline
    \end{tabular}
    \caption{Symmetry relations for the nonzero matrix elements of $\HH^I_{\mu,\bk}$ for the head-to-head domain wall. Each row contains four expressions that must, by symmetry, be equal.  The expressions are obtained by reflection in $x$ with $P_x\rightarrow -P_x$ ($\hat M_x$; second column), reflection in $y$ ($\hat M_y$; third column), and time reversal (TR; fourth column).  Note that Hermiticity requires $\epsilon^\alpha_\bk$ be real.}
    \label{Table:1}
\end{table}

\begin{table*}[tb]
    \begin{tabular}{c|c|c|c|c|c}
    Element & $\hat M_x$  & $\hat M_y$ & $\hat M_z$ & TR & $C_4$ \\
    \hline
        $\epsilon_{\bk}^\alpha$ & $\epsilon_{(-k_x,k_y)}^\alpha$ & $\epsilon_{(k_x,-k_y)}^\alpha$ & $\epsilon_{\bk}^\alpha$ & $(\epsilon_{-\bk}^\alpha)^*$ & $\epsilon^\alpha_{(k_y,-k_x)}$ \\
        $t_{0,\bk}^{xy,yz}$ & $-t_{0,(-k_x,k_y)}^{xy,yz}$ & $t_{0,(k_x,-k_y)}^{xy,yz}$ & $-t_{0,\bk}^{xy,yz}(-P_z)$ & $(t_{0,-\bk}^{xy,yz})^*$ & $-t_{0, (k_y,-k_x)}^{xy,zx}$\\
        $t_{0,\bk}^{xy,zx}$ & $t_{0,(-k_x,k_y)}^{xy,zx}$ & $-t_{0,(k_x,-k_y)}^{xy,zx}$ & $-t_{0,\bk}^{xy,zx}(-P_z)$ & $(t_{0,-\bk}^{xy,zx})^*$ & $t_{0,(k_y,-k_x)}^{xy,yz}$ \\
        $t_{0,\bk}^{yz,zx}$ & $-t_{0,(-k_x,k_y)}^{yz,zx}$ & $-t_{0,(k_x,-k_y)}^{yz,zx}$ & $t_{0,\bk}^{yz,zx}(-P_z)$ & $(t_{0,-\bk}^{yz,zx})^*$ & $-t_{0,(k_y,-k_x)}^{zx,yz}$\\
        \hline
        $t_{1,\bk}^{xy,yz}$ & $-t_{1,(-k_x,k_y)}^{xy,yz}$ & $-t_{1,(k_x,-k_y)}^{xy,yz}$ & $t_{1,\bk}^{xy,yz}(-P_z)$ & $(-t_{1,-\bk}^{xy,yz})^*$ & $-t_{2,(k_y,-k_x)}^{xy,zx}$ \\
        $t_{1,\bk}^{xy,zx}$ & $t_{1,(-k_x,k_y)}^{xy,zx}$ & $t_{1,(k_x,-k_y)}^{xy,zx}$ & $t_{1,\bk}^{xy,zx}(-P_z)$ & $(-t_{1,-\bk}^{xy,zx})^*$ & $t_{2, (k_y,-k_x)}^{xy,yz}$ \\
        $t_{1,\bk}^{yz,zx}$ & $-t_{1,(-k_x,k_y)}^{yz,zx}$ & $t_{1,(k_x,-k_y)}^{yz,zx}$ & $-t_{1,\bk}^{yz,zx}(-P_z)$ & $(-t_{1,-\bk}^{yz,zx})^*$ &  $-t_{2, (k_y,-k_x)}^{zx,yz}$ \\
        \hline
        $t_{2,\bk}^{xy,yz}$ & $t_{2,(-k_x,k_y)}^{xy,yz}$ & $t_{2,(k_x,-k_y)}^{xy,yz}$ & $t_{2,\bk}^{xy,yz}(-P_z)$ & $(-t_{2,-\bk}^{xy,yz})^*$ & $t_{1,(k_y,-k_x)}^{xy,zx}$ \\
        $t_{2,\bk}^{xy,zx}$ & $-t_{2,(-k_x,k_y)}^{xy,zx}$ & $-t_{2,(k_x,-k_y)}^{xy,zx}$ & $t_{2,\bk}^{xy,zx}(-P_z)$ & $(-t_{2,-\bk}^{xy,zx})^*$ & $-t_{1,(k_y,-k_x)}^{xy,yz}$ \\
        $t_{2,\bk}^{yz,zx}$ & $t_{2,(-k_x,k_y)}^{yz,zx}$ & $-t_{2,(k_x,-k_y)}^{yz,zx}$ & $-t_{2,\bk}^{yz,zx}(-P_z)$ & $(-t_{2,-\bk}^{yz,zx})^*$ & $t_{1,(k_y,-k_x)}^{zx,yz}$ \\
        \hline
        $t_{3,\bk}^{xy,yz}$ & $t_{3,(-k_x,k_y)}^{xy,yz}$ & $-t_{3,(k_x,-k_y)}^{xy,yz}$ & $-t_{3,\bk}^{xy,yz}(-P_z)$ & $(-t_{3,-\bk}^{xy,yz})^*$ & $-t_{3,(k_y,-k_x)}^{xy,zx}$\\
        $t_{3,\bk}^{xy,zx}$ & $-t_{3,(-k_x,k_y)}^{xy,zx}$ & $t_{3,(k_x,-k_y)}^{xy,zx}$ & $-t_{3,\bk}^{xy,zx}(-P_z)$ & $(-t_{3,-\bk}^{xy,zx})^*$ & $t_{3,(k_y,-k_x)}^{xy, yz}$\\
        $t_{3,\bk}^{yz,zx}$ & $t_{3,(-k_x,k_y)}^{yz,zx}$ & $t_{3,(k_x,-k_y)}^{xy,zx}$ & $t_{3,\bk}^{yz,zx}(-P_z)$ & $(-t_{3,-\bk}^{yz,zx})^*$ & $-t_{3,(k_y,-k_x)}^{zx,yz}$\\
        \hline
    \end{tabular}
    \caption{Symmetry relations for the nonzero matrix elements of $\HH^R_{\mu,\bk}$ for the head-to-tail domain wall. Each row contains six expressions that must, by symmetry, be equal.  The expressions are obtained by reflection in $x$, $y$, $z$ (with $P_z\rightarrow -P_z$),  time reversal, and $C_4$ rotation around $z$. }
    \label{Table:3}
\end{table*}

Our goal is to obtain the simplest tight-binding forms for the matrix elements of $\HH^I_{\mu,\bk}$ that are consistent with Table~\ref{Table:1}.
To begin, we take generic $t_{2g}$ tight-binding forms for the bare band dispersions that are applicable in two dimensions,
\begin{eqnarray}
    \epsilon^{xy}_\bk &=& - 2t_\| (c_x + c_y-2) + \epsilon^{xy}_0 \label{eq:exy} \\
    \epsilon^{yz}_\bk &=& - 2t_\| (c_y-1) - 2t_\perp (c_x-1) + \epsilon^{yz}_0 \label{eq:eyz} \\
    \epsilon^{zx}_\bk &=& - 2t_\| (c_x-1) - 2t_\perp (c_y-1) + \epsilon^{zx}_0 \label{eq:ezx}
\end{eqnarray}
with $c_{x,y} = \cos k_{x,y}$,
$t_\| = 200$~meV, $t_\perp = 35$~meV.  These values are typical for tight-binding models of BaTiO$_3$, as well as the quantum paraelectrics SrTiO$_3$ and KTaO$_3$.  The band offsets, $\epsilon_0^\alpha$, allow us to model different arrangements of the bands that can arise at domain walls.  
When $\epsilon_0^{xy}=\epsilon_0^{yz}=\epsilon_0^{zx}=0$, the bands are degenerate at the Brillouin zone centre. By comparison to density functional band structures \cite{salehi2003band}, we take $\epsilon_0^{yz} = -400 P_x^2$, with $P_x$ measured in C/m$^2$ to accommodate the tetragonal distortion of the unit cell by the polarization.  This polarization-dependent shift of $\epsilon_\bk^{yz}$ occurs in bulk ferroelectrics  and splits the degeneracy of the bands at $\bk=0$.

Next, we address the off-diagonal matrix elements.
To linear order in the polarization, the simplest expression for $t^{yz,zx}_{0,\bk}$ that is consistent with Table~\ref{Table:1} contains two terms,
\begin{equation}
    t^{yz,zx}_{0,\bk} = t_{01} s_x s_y + i t_{02} s_y P_x
\end{equation}
where $t_{01}$ and $t_{02}$ are real-valued tight-binding parameters and $s_x \equiv \sin k_x$, $s_y \equiv \sin k_y$.  The first term describes interorbital hopping processes between second-nearest-neighbour lattice sites, while  the second  describes the changes to the band structure due to inversion symmetry breaking by the polarization. The first of these terms has a quantitative effect on the band structure and is present whether or not the system is polarized, but does not contribute to the spin-orbit physics. We therefore set $t_{01}=0$ and focus on the role of inversion symmetry breaking.  By similar arguments, we keep only linear-in-polarization contributions to the remaining off-diagonal terms.  The complete list of inversion-breaking terms is
\begin{eqnarray}
    t_{0,\bk}^{yz,zx} &=& it_{0}^{yz,zx} s_y P_x \\
    t_{1,\bk}^{xy,yz} &=& t_{1}^{xy,yz} s_y P_x \\
    t_{1,\bk}^{xy,zx} &=& t_{1}^{xy,zx} s_x P_x \\
    t_{2,\bk}^{xy,yz} &=& t_{2}^{xy,yz} s_x P_x \\
    t_{2,\bk}^{xy,zx} &=& t_{2}^{xy,zx} s_y P_x \\
    t_{3,\bk}^{yz,zx} &=& t_{3}^{yz,zx} s_x P_x
\end{eqnarray}
Numerical diagonalization of $\HH_\bk$ reveals that terms proportional to $\sin k_y$ produce spin-splitting of the bands, while those proportional to $\sin k_x$ do not.  Unless otherwise stated, and to reduce the number of unknown parameters, we set $t_1^{xy,zx}=t_2^{xy,yz}=t_3^{yz,zx}=0$.

\subsection{Head-to-tail domain wall}

We next consider the head-to-tail domain wall shown in Fig.~\ref{fig:DWcartoon}(b).  Here, the polarization vector is perpendicular to wall, with ${\bf P} = P_z {\bf \hat z}$.
In addition to TR symmetry, the system has reflection symmetries $\hat M_x$, $\hat M_y$, and $\hat M_z$, where the latter includes the change of sign $P_z \rightarrow -P_z$.  To simplify the discussion, we further assume that there is a $C_4$ rotational symmetry around $z$.  The allowed nonzero matrix elements are given in Table~\ref{Table:3}.

Keeping only those off-diagonal terms that are either linear in $P_z$ or represent the atomic SOC, we get the Rashba-like Hamiltonian
\begin{eqnarray}
    \HH^R_{0,\bk} &=& \left [ \begin{array}{ccc}
         \epsilon^{xy}_\bk & ig_0 P_z s_x & ig_0 P_z s_y    \\
         -ig_0 P_z s_x & \epsilon^{yz}_\bk & 0 \\
         -ig_0 P_z s_y& 0 & \epsilon^{zx}_\bk 
    \end{array}\right ], \\
    \HH^R_{1,\bk} &=& \left [ \begin{array}{ccc}
         0 & 0 & -i\frac \xi 2 \\
         0 &0& g_1  P_z s_x \\
         i\frac \xi 2 & g_1  P_z s_x  & 0
    \end{array}\right ], \\
    \HH^R_{2,\bk} &=& \left [ \begin{array}{ccc}
         0 & i\frac \xi 2 & 0 \\
         -i\frac \xi 2 &0& g_1  P_z s_y  \\
         0 & g_1 P_z s_y & 0
    \end{array}\right ], \\
    \HH^R_{3,\bk} &=& \left [ \begin{array}{ccc}
         0 & g_3  P_z s_y & -g_3 P_z s_x \\
         g_3 P_z s_y &0& i \frac\xi 2 \\
         -g_3 P_z s_x &-i \frac\xi 2& 0
    \end{array}\right ].
\end{eqnarray}
The bare dispersions are, as before, given by Eqs.~(\ref{eq:exy})-(\ref{eq:ezx}), but with $\epsilon_0^{yz}=\epsilon_0^{zx}=0$ and $\epsilon_0^{xy}=-400P_z^2$.  It is usual to make the further approximation that $g_1=g_3=0$ \cite{zhong2013theory}; these additional terms affect the $k$-dependent amplitude of the spin-splitting around the Fermi surface and should, in general, be included.  However, to facilitate comparisons to previous work we set them to zero here.

\section{Results}
\label{sec:results}

\subsection{Band structure}
\begin{figure}[tb]
    \includegraphics[width=\columnwidth]{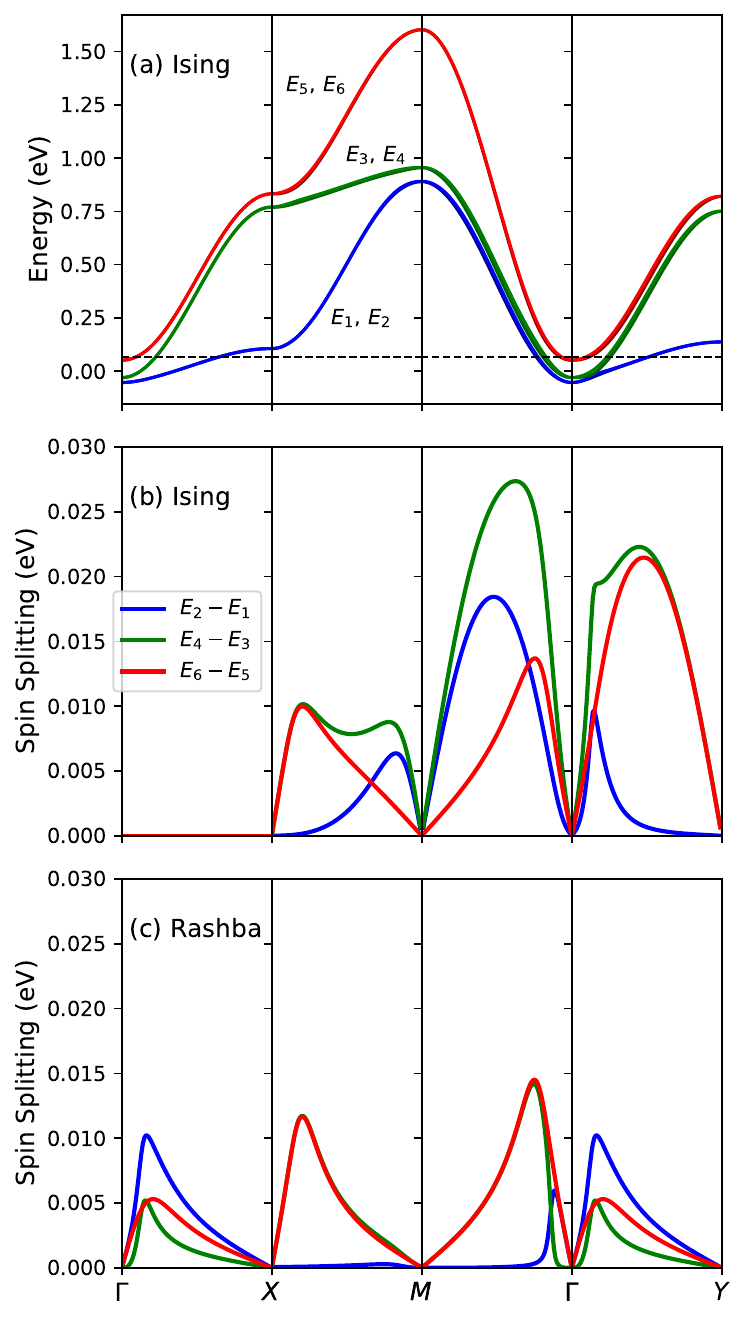}
    \caption{Band structure of the six-band model.  (a) Band dispersion and (b) spin splitting for a head-to-head domain wall with Ising SOC; (c) spin-splitting for a head-to-tail domain wall with Rashba SOC.  The horizontal dashed line in (a) indicates the energy of the contours in Fig.~\ref{fig:Ising-FS}. Results are for $\xi=60$~meV and (a),(b) $P_x=0.28$~C/m$^2$ and $t_0^{yz,zx} =t_1^{xy,yz}=t_2^{xy,zx}=40$~meV$\cdot\mathrm{m}^2/\mathrm{C}$ or (c) $P_z=0.28$~C/m$^2$ and $g_0 = 40~\mathrm{meV}\cdot\mathrm{m}^2/\mathrm{C}$.  Labels $E_i$ indicate the different bands, and colours have the same meaning in (b) and (c).}
    \label{fig:Ising-bands}
\end{figure}

\begin{figure}[tb]
    \includegraphics[width=\columnwidth]{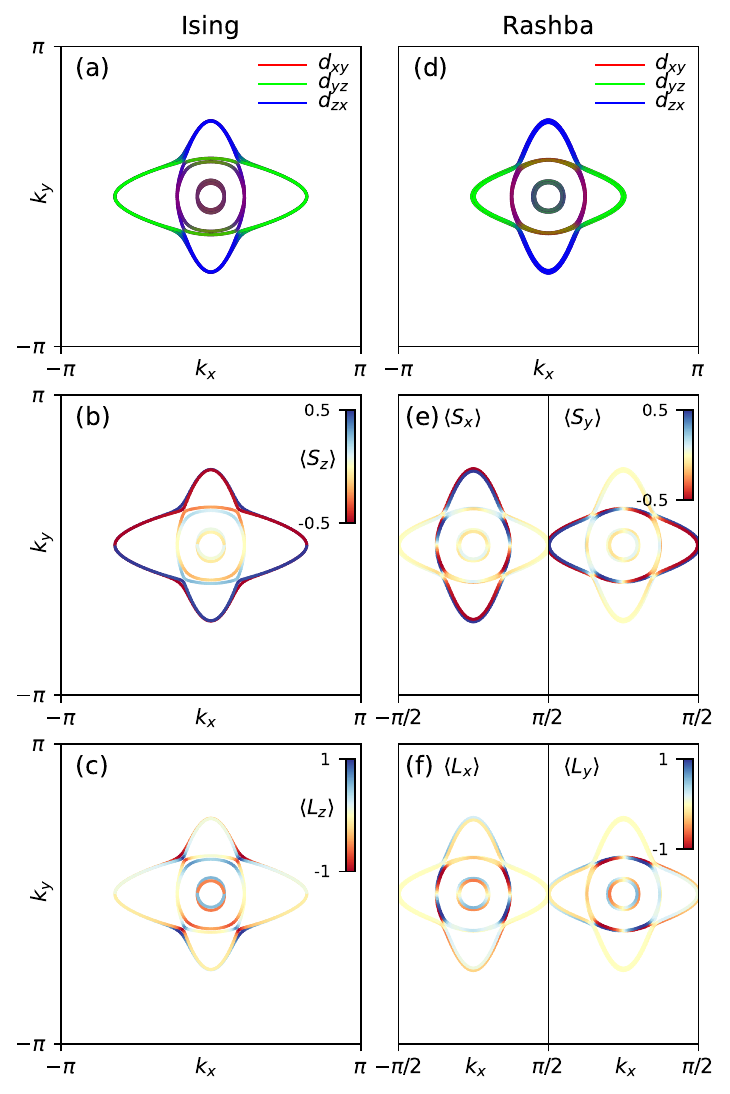}
    \caption{Orbital and spin properties of the (a)-(c) Ising and (d)-(f) Rashba Hamiltonians.  The orbital character (a), (d); nonzero components of the electron spin (b), (e); and orbital angular momenta (c), (f) are shown. Parameters are as in Fig.~\ref{fig:Ising-bands}. }
    \label{fig:Ising-FS}
\end{figure}

We show the effects of SOC on the band structure in Figs.~\ref{fig:Ising-bands} and \ref{fig:Ising-FS} for both the Ising and Rashba cases.  For illustrative purposes we take spin-orbit parameters that are large enough that the effects of SOC are plainly visible. The atomic SOC, $\xi = 60$~meV, is three times its value in Ti (20~meV \cite{TF9615701441,zhong2013theory}), is comparable to its value in Nb (80~meV \cite{TF9615701441}), and is one-fifth of its value in Ta (300~meV \cite{Ishikawa:2019ASOC}); the Ising- and Rashba-specific parameters are not known for domain walls; however, $g_0P_z \approx 5$~meV has been proposed for SrTiO$_3$ interfaces \cite{vivek2017topological}.  Our choices of $g_0=40$~meV$\cdot$m$^2$/C and $P_z = 0.28$~C/m$^2$, give approximately twice this.  Since we have no guidance on the Ising-specific parameters, we also take $t_0^{yz,zx} =t_1^{xy,yz}=t_2^{xy,zx}=40$~meV$\cdot\mathrm{m}^2/\mathrm{C}$. 

Figure~\ref{fig:Ising-bands}(a) shows the band structure along high-symmetry directions in the Brillouin zone for the Ising-SOC model (the Rashba model is nearly identical).  Although six bands are plotted, only three curves can be distinguished easily because SOC is a weak perturbation.  Figures~\ref{fig:Ising-bands}(b) and (c) show the so-called spin-splitting of the bands, namely the energy difference between closely spaced pairs of bands with the same orbital character but opposite spin textures.   For both models, the spin-splitting depends on both the amplitudes and signs of the inversion-breaking terms---$t_0^{yz,zx}$, $t_1^{xy,yz}$, and $t_2^{xy,zx}$ or $g_0$, $g_1$, and $g_3$---but are quantitatively similar when the terms are comparable.  The qualitative difference between the two models is that the Rashba spin-splitting has fourfold symmetry (it is the same along $\Gamma$-$X$ and $\Gamma$-$Y$), while the Ising spin-splitting vanishes when $k_y=0$.  

Because measured electron densities at domain walls are low, we focus on the band structure near $\Gamma$. At domain walls, the band degeneracy at $\Gamma$ is broken by three factors.  First, as discussed above, the contributions to $\epsilon_0^\alpha$ in Eqs.~(\ref{eq:exy})-(\ref{eq:ezx}) that are proportional to $P_x^2$ or $P_z^2$ describe the symmetry breaking of the bulk bands by the polarization. For the Ising case,  this term creates an orthorhombic distortion of the energy contours shown in Figs.~\ref{fig:Ising-FS}(a)-(c). Tetragonal symmetry is preserved, however, for the Rashba case shown in  Figs.~\ref{fig:Ising-FS}(d)-(f). Second, a rigid band shift of $\epsilon_0^{xy}$ is expected because the domain wall confines the bands to a narrow channel in the xy plane. This effect is well-known at SrTiO$_3$ interfaces, where bands that are heavy perpendicular to the surface are lower in energy than those that are light (see, for example, \cite{Raslan:2017temperature}).  In Ref.~\cite{cornell2023influence}, however, it was shown that this effect is much smaller at domain walls, and for this reason we ignore it here.  Third, the atomic SOC lifts the band degeneracy at $\Gamma$ \cite{zhong2013theory}.

The dashed horizontal line in Fig.~\ref{fig:Ising-bands}(a) indicates the energy of the constant-energy-contours shown in Fig.~\ref{fig:Ising-FS}.   If we take this to be the Fermi energy, then it corresponds to a filling of $0.4$~e$^-$ per unit cell.  For reference, this is comparable to what is required to compensate the bound charge at the domain wall in BaTiO$_3$: for a typical value of $P=0.3$~C/m$^2$ in the bulk phases on either side of 90$^\circ$ domain wall,  $({\bf P}_1-{\bf P}_2)\cdot {\bf \hat n} = \sqrt{2} P \approx 0.4$~C/m$^2$. (For a lattice constant of 4~\AA, this is equal to 0.4~e$^-$ per u.c.) In practice, observed electron densities are far below this value.

Figure \ref{fig:Ising-FS} compares the orbital texture, spin, and orbital angular momentum of the Ising and Rashba Hamiltonians.  For the Ising case, both the spin and orbital angular momentum are entirely polarized along the $z$ axis, perpendicular to the domain wall.  Conversely, for the Rashba case shown in Figs.~\ref{fig:Ising-FS}(d)-(f), the spin and orbital angular momenta lie entirely in the plane of the wall, and there is an obvious connection between the orbital character and the direction of the angular momentum  (sections of contour with $d_{yz}$ character, for example, have angular momentum along $y$).  Apart from their distinct symmetries, the most striking difference between the Rashba and Ising cases is that, while the orbital angular momentum is distributed across much of the Fermi surface in the Rashba case, it is concentrated near the avoided band crossings in the Ising case.

\subsection{Edelstein effect}
\begin{figure}[tb]
    \centering
    \includegraphics[width=\columnwidth]{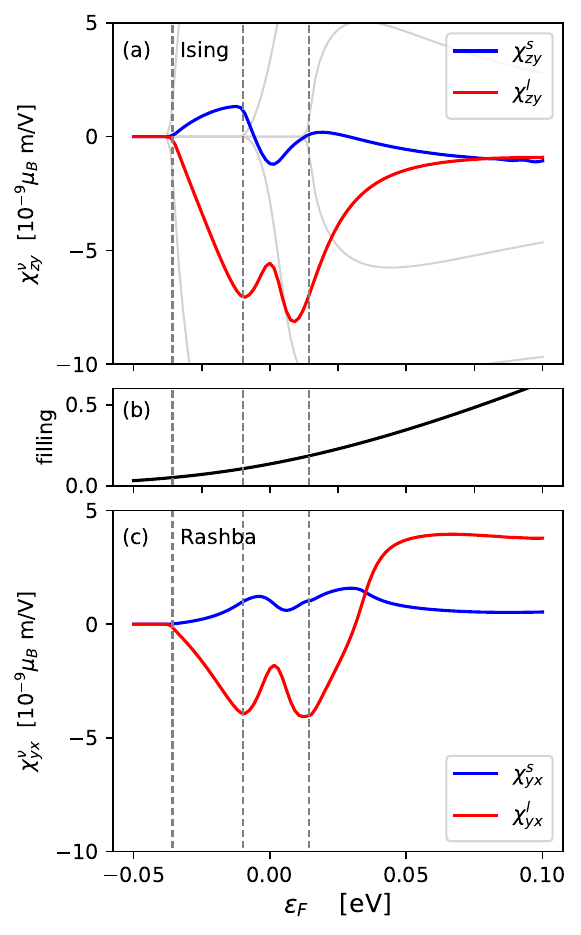}
    \caption{Spin (blue) and orbital (red) magnetization conversion efficiencies as functions of the the Fermi energy $\varepsilon_F$ for (a) the Ising and (c) the Rashba cases in the degenerate limit ($T\rightarrow 0$).  Panel (b) shows the electron filling per unit cell as a function of $\varepsilon_F$.  Vertical dashed lines indicate energies of the band bottoms (note that their spin-splitting is too small to resolve).  Light grey lines in (a) show the individual band contributions to the orbital conversion efficiency.   Here,  $\xi=20$~meV and (a) $t_0^{yz,zx} =t_1^{xy,yz}=t_2^{xy,zx}=15$~meV$\cdot\mathrm{m}^2/\mathrm{C}$ or (b) $g_0 = 15~\mathrm{meV}\cdot\mathrm{m}^2/\mathrm{C}$.}
    \label{fig:ising-edelstein}
\end{figure}

Observable consequences of these angular momentum textures can be found in the Edelstein effect, which describes the magnetization per unit cell ${\bf m}$  that is induced by an electric field ${\bf E}$.  The magnetization has both spin and orbital contributions, and in the linear regime one writes
\begin{equation}
    m_i = \sum_j (\chi^s_{ij} + \chi^l_{ij}) E_j 
\end{equation}
where $i,j \in (x,y,z)$ are spatial directions and $\chi^s_{ij}$ and $\chi^l_{ij}$ are the spin and orbital conversion efficiencies.  For the head-to-head domain wall, both the spin and orbital magnetization are parallel to the $z$ axis,  and are only nonzero when the electric field is parallel to the $y$ direction.
Taking the semiclassical equations from Johansson \textit{et al.} \cite{johansson2021spin}, we have
\begin{eqnarray}
  \chi^s_{ij} &=& \frac{e g_s \mu_B\tau_0 }{\hbar}  \frac{1}{N_k} \sum_\bk \sum_\nu s^i_{\nu\bk}  v^j_{\nu\bk} \frac{\partial f(E_{\nu\bk})}{\partial E_{\nu\bk}},\\
    \chi^l_{ij} &=& \frac{e\mu_B\tau_0 }{\hbar}  \frac{1}{N_k} \sum_\bk \sum_\nu  L^i_{\nu\bk}  v^j_{\nu\bk} \frac{\partial f(E_{\nu\bk})}{\partial E_{\nu\bk}},
\end{eqnarray}
where $E_{\nu\bk}$ is the dispersion for band $\nu$, $v^j_{\nu\bk} = \frac 1\hbar \partial E_{\nu\bk}/\partial k_j$ is the corresponding quasiparticle velocity,  $\tau_0$ is the transport lifetime (taken to be 1~ps), $g_s \approx 2$ is the Land\'e $g$-factor for electrons, $\mu_B$ is the Bohr magneton.   The spin matrix elements are  $s^i_{\nu \bk} = \frac 12 \langle \nu \bk| [1_{3\times3}\otimes \sigma_{i} ]| \nu \bk \rangle$, where $|\nu \bk\rangle$ are eigenstates of the Hamiltonian with band energies $E_{\nu\bk}$.  The orbital angular momentum is, to a good approximation \cite{johansson2021spin}, given by the intra-atomic contribution, $L^i_{\nu\bk} = \langle \nu \bk| [l_i\otimes \sigma_{0} ]| \nu \bk \rangle$, with 
\begin{equation}
l_x =\left[ \begin{array}{ccc}
     0 & 0 & -i  \\
     0 & 0 & 0 \\
     i & 0 & 0 
\end{array}\right], \,
l_y = \left[ \begin{array}{ccc}
0 & i & 0 \\
-i & 0 & 0 \\
0 & 0 & 0 
\end{array}\right], \,
l_z = \left[ \begin{array}{ccc}
0 & 0 & 0 \\
0 & 0 & i \\
0 & -i & 0 
\end{array}\right].
\end{equation}

The conversion efficiencies can be shown analytically to vanish for the two-band Ising Hamiltonian introduced in Sec.~\ref{sec:twoband}:
$\chi^l_{zy}$ vanishes trivially because the orbital angular momentum of the underlying Wannier orbitals is assumed to vanish; $\chi^s_{zy}$ vanishes because the spin $s^z_{\nu\bk}$ is constant in each band while the Fermi velocity given by Eq.~(\ref{eq:HI1bd}) integrates to zero around each circular Fermi surface.

The situation is different in the multiorbital $t_{2g}$ model, as shown in Fig.~\ref{fig:ising-edelstein}(a).  Here, we have taken spin-orbit parameters that are motivated by SrTiO$_3$ interfaces \cite{zhong2013theory}.  In this case, the $k$-dependent orbital mixing generates nonzero spin and orbital conversion efficiencies, with the orbital efficiency nearly an order of magnitude larger than the spin efficiency.  This is consistent with previous observations that for  SrTiO$_3$ interfaces with Rashba SOC, the orbital conversion efficiency is roughly an order of magnitude larger than the spin conversion efficiency \cite{johansson2021spin,el2023observation}. Indeed, the conversion efficiencies  shown in Fig.~\ref{fig:ising-edelstein}(c) for the Rashba model show a similar disparity between spin and orbital contributions.  The current-induced magnetisation at CDWs, therefore, is predicted to be predominantly due to the orbital Edelstein effect.

\section{Discussion}
The prospect of obtaining robust spin-orbit physics at CDWs is enticing because of the potential for reconfigurable circuits with spin-orbitronic functionalities.  The key question is whether such an outcome is practically feasible.  Of the three most widely studied materials hosting CDWs, LiNbO$_3$ has the highest reported electron mobilities ($> 1000$~cm$^2$/Vs \cite{McClusky:2025}) but hosts 180$^\circ$ head-to-head walls that do not appear to break inversion symmetry.  Smaller, but still large, mobilities ($>100$~cm$^2$/Vs \cite{Beccard:2022}) are reported for 90$^\circ$ head-to-head domain walls in tetragonal BaTiO$_3$, which appears therefore to be a strong candidate in which to observe the proposed Ising-type spin-orbit physics.  Mobilities are not reported for head-to-tail walls in BiFeO$_3$; however, reported conductances are quite small ($\sim$pA/V) \cite{zhang2019intrinsic}, which likely  makes direct observation of the expected Rashba-type spin-orbit physics challenging.   

\begin{figure}
    \centering
    \includegraphics[width=\columnwidth]{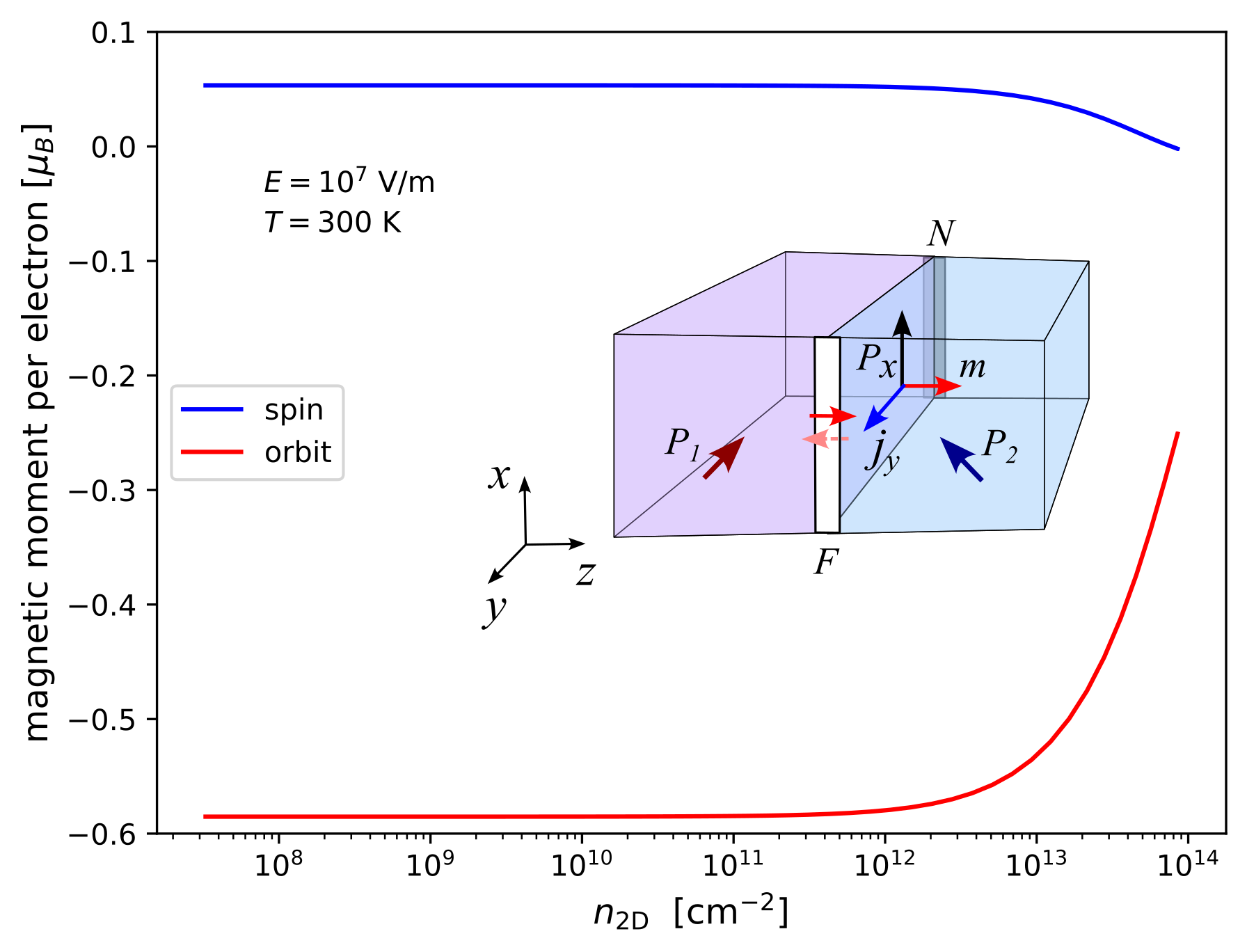}
    \caption{Magnetic moment per electron induced by a typical electric field, $E=10^7$~V/m at $T=300$~K as a function of electron density.  Other parameters are as in Fig.~\ref{fig:ising-edelstein}.  The inset shows a simple experiment to measure the magnetization. The ferroelectric polarizations ${\bf P}_1$ and ${\bf P}_2$ in each domain result in a N\'eel polarization $P_x$ along the domain wall.  A current density $j_y$ generates a magnetization $m$ perpendicular to the wall. The current is driven between a normal metal lead (N) and a ferromagnetic (F) lead that is magnetized parallel (solid arrow) or anti-parallel (dashed arrow) to $m$.}
    \label{fig:edelstein}
\end{figure}

This suggests that BaTiO$_3$ is the best candidate for demonstrating spin-orbit physics.  An important caveat, however, is that  the band description of the electron gas at the domain wall is not established.  Generally, domain-wall conductivity measurements observe an activated (Arrhenius) temperature dependence,  $\sigma(T) = \sigma_0 \exp(-E_a/k_BT)$ with an activation energy $E_a\sim 0.1$-$0.2$~eV.  This temperature dependence has been attributed variously to  thermally assisted hopping \cite{Zahn:2024} and to thermal excitation of electrons from shallow donors into the conduction band \cite{seidel2010domain}.  If the latter holds, then the band picture makes sense.  

Making this assumption, we show in Fig.~\ref{fig:edelstein} the spin and orbital magnetic moments induced by an electric field for parameters appropriate to BaTiO$_3$.  In particular, we show results for low electron densities and at room temperature where head-to-head $90^\circ$ domain walls are obeserved.  Because of the extreme low densities in experiments (as low as $10^3$~cm$^{-2}$ \cite{Beccard:2022}), absolute magnetizations are  immeasurably small; however, as shown in Fig.~\ref{fig:edelstein}, a typical electric field of  $10^7$~V/m induces a substantial magnetic moment \textit{per electron}.  Because of this, the (small) current induced by this field should have significant spin and orbital polarizations.  

The inset to Fig.~\ref{fig:edelstein} shows a conceptually simple transport experiment that is loosely based on a spin-injection technique used to demonstrate spin-charge conversion in InAs quantum wells \cite{hammar2000potentiometric}.  In Fig.~\ref{fig:edelstein}, a 90$^\circ$ domain wall separating regions with ferroelectric polarizations ${\bf P}_1$ and ${\bf P}_2$ extends the length of a ferroelectric mesa.   The polarization ${P}_x$ along the wall is taken to be N\'eel-type, and points upwards.  Normal metal (N) and ferromagnetic (F) leads are attached to opposite ends of the wall in such a way that a charge current $j_y$ can be driven along the axis orthogonal to ${P}_x$. (In practice, the leads may also be top-mounted, with one being a conducting AFM tip.) The current induces a magnetization $m$ in the 2DEG along the positive or negative $z$ axis (perpendicular to the wall) depending on the sign of ${j}_y$.  The ferromagnet's magnetization ${m}_F$ can be switched between the $+z$ and $-z$ directions by an external magnetic field.  The key point of this experiment is that the junction resistance between the wall and drain depends on the relative directions of $m$ and ${m}_F$ \cite{hammar2000potentiometric}. The experimental signature of spin-charge conversion is that the junction  switches between low- and high-resistance states when either $m_F$ or $j_y$ changes sign.

\section{Conclusions}
We have explored the role of spin-orbit physics at conducting ferroelectric domain walls with broken inversion symmetry.  These include both head-to-head and head-to-tail walls with either N\'eel or Bloch polarization profiles. In the former case, the $k$-space spin and angular momentum textures are Ising-like, with magnetization perpendicular to the wall; in the latter case, the textures are Rashba-like, with magnetization parallel to the wall.  We have argued that even at the very low electron densities that are typical of conducting domain walls, the electron gas may be magnetized by an electric field via the Edelstein effect, and that this  should be observable experimentally through spin-injection experiments.

\section*{Acknowledgments}
We acknowledge the support of the Natural Sciences and Engineering Research Council of Canada
(NSERC). 


\begin{thebibliography}{65}%
\makeatletter
\providecommand \@ifxundefined [1]{%
 \@ifx{#1\undefined}
}%
\providecommand \@ifnum [1]{%
 \ifnum #1\expandafter \@firstoftwo
 \else \expandafter \@secondoftwo
 \fi
}%
\providecommand \@ifx [1]{%
 \ifx #1\expandafter \@firstoftwo
 \else \expandafter \@secondoftwo
 \fi
}%
\providecommand \natexlab [1]{#1}%
\providecommand \enquote  [1]{``#1''}%
\providecommand \bibnamefont  [1]{#1}%
\providecommand \bibfnamefont [1]{#1}%
\providecommand \citenamefont [1]{#1}%
\providecommand \href@noop [0]{\@secondoftwo}%
\providecommand \href [0]{\begingroup \@sanitize@url \@href}%
\providecommand \@href[1]{\@@startlink{#1}\@@href}%
\providecommand \@@href[1]{\endgroup#1\@@endlink}%
\providecommand \@sanitize@url [0]{\catcode `\\12\catcode `\$12\catcode
  `\&12\catcode `\#12\catcode `\^12\catcode `\_12\catcode `\%12\relax}%
\providecommand \@@startlink[1]{}%
\providecommand \@@endlink[0]{}%
\providecommand \url  [0]{\begingroup\@sanitize@url \@url }%
\providecommand \@url [1]{\endgroup\@href {#1}{\urlprefix }}%
\providecommand \urlprefix  [0]{URL }%
\providecommand \Eprint [0]{\href }%
\providecommand \doibase [0]{https://doi.org/}%
\providecommand \selectlanguage [0]{\@gobble}%
\providecommand \bibinfo  [0]{\@secondoftwo}%
\providecommand \bibfield  [0]{\@secondoftwo}%
\providecommand \translation [1]{[#1]}%
\providecommand \BibitemOpen [0]{}%
\providecommand \bibitemStop [0]{}%
\providecommand \bibitemNoStop [0]{.\EOS\space}%
\providecommand \EOS [0]{\spacefactor3000\relax}%
\providecommand \BibitemShut  [1]{\csname bibitem#1\endcsname}%
\let\auto@bib@innerbib\@empty
\bibitem [{\citenamefont {Guro}\ \emph {et~al.}(1968)\citenamefont {Guro},
  \citenamefont {Ivanchik},\ and\ \citenamefont {Kovtonyuk}}]{Guro:1968}%
  \BibitemOpen
  \bibfield  {author} {\bibinfo {author} {\bibfnamefont {G.~M.}\ \bibnamefont
  {Guro}}, \bibinfo {author} {\bibfnamefont {I.~I.}\ \bibnamefont {Ivanchik}},\
  and\ \bibinfo {author} {\bibfnamefont {N.~F.}\ \bibnamefont {Kovtonyuk}},\
  }\bibfield  {title} {\bibinfo {title} {Semiconductor {Properties} of {Barium}
  {Titanate}},\ }\href@noop {} {\bibfield  {journal} {\bibinfo  {journal}
  {Soviet Physics - Solid State}\ }\textbf {\bibinfo {volume} {10}},\ \bibinfo
  {pages} {100} (\bibinfo {year} {1968})},\ \bibinfo {note} {[Fiz. Tverd. Tela
  {\bf 10}, 135 (1968)]}\BibitemShut {NoStop}%
\bibitem [{\citenamefont {{G M Guro, I I Ivanchik, N F
  Kovtonyuk}}(1970)}]{Guro:1970}%
  \BibitemOpen
  \bibfield  {author} {\bibinfo {author} {\bibnamefont {{G M Guro, I I
  Ivanchik, N F Kovtonyuk}}},\ }\bibfield  {title} {\bibinfo {title}
  {c-{Domain} {Barium} {Titanate} {Crystal} in a {Short}-{Circuited}
  {Capacitor}},\ }\href@noop {} {\bibfield  {journal} {\bibinfo  {journal}
  {Soviet Physics - Solid State}\ }\textbf {\bibinfo {volume} {11}},\ \bibinfo
  {pages} {1574} (\bibinfo {year} {1970})}\BibitemShut {NoStop}%
\bibitem [{\citenamefont {Krapivin}\ and\ \citenamefont
  {Chenskii}(1970)}]{Krapivin:1970}%
  \BibitemOpen
  \bibfield  {author} {\bibinfo {author} {\bibfnamefont {V.~F.}\ \bibnamefont
  {Krapivin}}\ and\ \bibinfo {author} {\bibfnamefont {E.~V.}\ \bibnamefont
  {Chenskii}},\ }\bibfield  {title} {\bibinfo {title} {Space-{Charge}-{Limited}
  {Currents} in a {Metal}-{Ferroelectric}-{Metal} {System}},\ }\href@noop {}
  {\bibfield  {journal} {\bibinfo  {journal} {Soviet Physics - Solid State}\
  }\textbf {\bibinfo {volume} {12}},\ \bibinfo {pages} {454} (\bibinfo {year}
  {1970})},\ \bibinfo {note} {[Fiz. Tverd. Tela {\bf 12}, 597
  (1970)]}\BibitemShut {NoStop}%
\bibitem [{\citenamefont {Vul}\ \emph {et~al.}(1973)\citenamefont {Vul},
  \citenamefont {Guro},\ and\ \citenamefont {Ivanchik}}]{Vul:1973}%
  \BibitemOpen
  \bibfield  {author} {\bibinfo {author} {\bibfnamefont {B.~M.}\ \bibnamefont
  {Vul}}, \bibinfo {author} {\bibfnamefont {G.~M.}\ \bibnamefont {Guro}},\ and\
  \bibinfo {author} {\bibfnamefont {I.~I.}\ \bibnamefont {Ivanchik}},\
  }\bibfield  {title} {\bibinfo {title} {Encountering domains in
  ferroelectrics},\ }\href {https://doi.org/10.1080/00150197308237691}
  {\bibfield  {journal} {\bibinfo  {journal} {Ferroelectrics}\ }\textbf
  {\bibinfo {volume} {6}},\ \bibinfo {pages} {29} (\bibinfo {year}
  {1973})}\BibitemShut {NoStop}%
\bibitem [{\citenamefont {Seidel}\ \emph {et~al.}(2009)\citenamefont {Seidel},
  \citenamefont {Martin}, \citenamefont {He}, \citenamefont {Zhan},
  \citenamefont {Chu}, \citenamefont {Rother}, \citenamefont {Hawkridge},
  \citenamefont {Maksymovych}, \citenamefont {Yu}, \citenamefont {Gajek} \emph
  {et~al.}}]{seidel2009conduction}%
  \BibitemOpen
  \bibfield  {author} {\bibinfo {author} {\bibfnamefont {J.}~\bibnamefont
  {Seidel}}, \bibinfo {author} {\bibfnamefont {L.~W.}\ \bibnamefont {Martin}},
  \bibinfo {author} {\bibfnamefont {Q.}~\bibnamefont {He}}, \bibinfo {author}
  {\bibfnamefont {Q.}~\bibnamefont {Zhan}}, \bibinfo {author} {\bibfnamefont
  {Y.-H.}\ \bibnamefont {Chu}}, \bibinfo {author} {\bibfnamefont
  {A.}~\bibnamefont {Rother}}, \bibinfo {author} {\bibfnamefont
  {M.}~\bibnamefont {Hawkridge}}, \bibinfo {author} {\bibfnamefont
  {P.}~\bibnamefont {Maksymovych}}, \bibinfo {author} {\bibfnamefont
  {P.}~\bibnamefont {Yu}}, \bibinfo {author} {\bibfnamefont {M.}~\bibnamefont
  {Gajek}}, \emph {et~al.},\ }\bibfield  {title} {\bibinfo {title} {Conduction
  at domain walls in oxide multiferroics},\ }\href@noop {} {\bibfield
  {journal} {\bibinfo  {journal} {Nature materials}\ }\textbf {\bibinfo
  {volume} {8}},\ \bibinfo {pages} {229} (\bibinfo {year} {2009})}\BibitemShut
  {NoStop}%
\bibitem [{\citenamefont {Seidel}\ \emph {et~al.}(2010)\citenamefont {Seidel},
  \citenamefont {Maksymovych}, \citenamefont {Batra}, \citenamefont {Katan},
  \citenamefont {Yang}, \citenamefont {He}, \citenamefont {Baddorf},
  \citenamefont {Kalinin}, \citenamefont {Yang}, \citenamefont {Yang} \emph
  {et~al.}}]{seidel2010domain}%
  \BibitemOpen
  \bibfield  {author} {\bibinfo {author} {\bibfnamefont {J.}~\bibnamefont
  {Seidel}}, \bibinfo {author} {\bibfnamefont {P.}~\bibnamefont {Maksymovych}},
  \bibinfo {author} {\bibfnamefont {Y.}~\bibnamefont {Batra}}, \bibinfo
  {author} {\bibfnamefont {A.}~\bibnamefont {Katan}}, \bibinfo {author}
  {\bibfnamefont {S.-Y.}\ \bibnamefont {Yang}}, \bibinfo {author}
  {\bibfnamefont {Q.}~\bibnamefont {He}}, \bibinfo {author} {\bibfnamefont
  {A.~P.}\ \bibnamefont {Baddorf}}, \bibinfo {author} {\bibfnamefont {S.~V.}\
  \bibnamefont {Kalinin}}, \bibinfo {author} {\bibfnamefont {C.-H.}\
  \bibnamefont {Yang}}, \bibinfo {author} {\bibfnamefont {J.-C.}\ \bibnamefont
  {Yang}}, \emph {et~al.},\ }\bibfield  {title} {\bibinfo {title} {Domain wall
  conductivity in {La}-doped {BiFeO$_3$}},\ }\href@noop {} {\bibfield
  {journal} {\bibinfo  {journal} {Phys. Rev. Lett.}\ }\textbf {\bibinfo
  {volume} {105}},\ \bibinfo {pages} {197603} (\bibinfo {year}
  {2010})}\BibitemShut {NoStop}%
\bibitem [{\citenamefont {Maksymovych}\ \emph {et~al.}(2011)\citenamefont
  {Maksymovych}, \citenamefont {Seidel}, \citenamefont {Chu}, \citenamefont
  {Wu}, \citenamefont {Baddorf}, \citenamefont {Chen}, \citenamefont
  {Kalinin},\ and\ \citenamefont {Ramesh}}]{maksymovych2011dynamic}%
  \BibitemOpen
  \bibfield  {author} {\bibinfo {author} {\bibfnamefont {P.}~\bibnamefont
  {Maksymovych}}, \bibinfo {author} {\bibfnamefont {J.}~\bibnamefont {Seidel}},
  \bibinfo {author} {\bibfnamefont {Y.~H.}\ \bibnamefont {Chu}}, \bibinfo
  {author} {\bibfnamefont {P.}~\bibnamefont {Wu}}, \bibinfo {author}
  {\bibfnamefont {A.~P.}\ \bibnamefont {Baddorf}}, \bibinfo {author}
  {\bibfnamefont {L.-Q.}\ \bibnamefont {Chen}}, \bibinfo {author}
  {\bibfnamefont {S.~V.}\ \bibnamefont {Kalinin}},\ and\ \bibinfo {author}
  {\bibfnamefont {R.}~\bibnamefont {Ramesh}},\ }\bibfield  {title} {\bibinfo
  {title} {Dynamic conductivity of ferroelectric domain walls in {BiFeO$_3$}},\
  }\href@noop {} {\bibfield  {journal} {\bibinfo  {journal} {Nano letters}\
  }\textbf {\bibinfo {volume} {11}},\ \bibinfo {pages} {1906} (\bibinfo {year}
  {2011})}\BibitemShut {NoStop}%
\bibitem [{\citenamefont {Farokhipoor}\ and\ \citenamefont
  {Noheda}(2011)}]{farokhipoor2011conduction}%
  \BibitemOpen
  \bibfield  {author} {\bibinfo {author} {\bibfnamefont {S.}~\bibnamefont
  {Farokhipoor}}\ and\ \bibinfo {author} {\bibfnamefont {B.}~\bibnamefont
  {Noheda}},\ }\bibfield  {title} {\bibinfo {title} {Conduction through
  71$^\circ$ domain walls in {BiFeO$_3$} thin films},\ }\href@noop {}
  {\bibfield  {journal} {\bibinfo  {journal} {Phys. Rev. Lett}\ }\textbf
  {\bibinfo {volume} {107}},\ \bibinfo {pages} {127601} (\bibinfo {year}
  {2011})}\BibitemShut {NoStop}%
\bibitem [{\citenamefont {Rojac}\ \emph {et~al.}(2017)\citenamefont {Rojac},
  \citenamefont {Bencan}, \citenamefont {Drazic}, \citenamefont {Sakamoto},
  \citenamefont {Ursic}, \citenamefont {Jancar}, \citenamefont {Tavcar},
  \citenamefont {Makarovic}, \citenamefont {Walker}, \citenamefont {Malic}
  \emph {et~al.}}]{rojac2017domain}%
  \BibitemOpen
  \bibfield  {author} {\bibinfo {author} {\bibfnamefont {T.}~\bibnamefont
  {Rojac}}, \bibinfo {author} {\bibfnamefont {A.}~\bibnamefont {Bencan}},
  \bibinfo {author} {\bibfnamefont {G.}~\bibnamefont {Drazic}}, \bibinfo
  {author} {\bibfnamefont {N.}~\bibnamefont {Sakamoto}}, \bibinfo {author}
  {\bibfnamefont {H.}~\bibnamefont {Ursic}}, \bibinfo {author} {\bibfnamefont
  {B.}~\bibnamefont {Jancar}}, \bibinfo {author} {\bibfnamefont
  {G.}~\bibnamefont {Tavcar}}, \bibinfo {author} {\bibfnamefont
  {M.}~\bibnamefont {Makarovic}}, \bibinfo {author} {\bibfnamefont
  {J.}~\bibnamefont {Walker}}, \bibinfo {author} {\bibfnamefont
  {B.}~\bibnamefont {Malic}}, \emph {et~al.},\ }\bibfield  {title} {\bibinfo
  {title} {Domain-wall conduction in ferroelectric {BiFeO$_3$} controlled by
  accumulation of charged defects},\ }\href@noop {} {\bibfield  {journal}
  {\bibinfo  {journal} {Nature materials}\ }\textbf {\bibinfo {volume} {16}},\
  \bibinfo {pages} {322} (\bibinfo {year} {2017})}\BibitemShut {NoStop}%
\bibitem [{\citenamefont {Zhang}\ \emph
  {et~al.}(2019{\natexlab{a}})\citenamefont {Zhang}, \citenamefont {Lu},
  \citenamefont {Yan}, \citenamefont {Cheng}, \citenamefont {Xie},
  \citenamefont {Aoki}, \citenamefont {Li}, \citenamefont {Heikes},
  \citenamefont {Lau}, \citenamefont {Schlom} \emph
  {et~al.}}]{zhang2019intrinsic}%
  \BibitemOpen
  \bibfield  {author} {\bibinfo {author} {\bibfnamefont {Y.}~\bibnamefont
  {Zhang}}, \bibinfo {author} {\bibfnamefont {H.}~\bibnamefont {Lu}}, \bibinfo
  {author} {\bibfnamefont {X.}~\bibnamefont {Yan}}, \bibinfo {author}
  {\bibfnamefont {X.}~\bibnamefont {Cheng}}, \bibinfo {author} {\bibfnamefont
  {L.}~\bibnamefont {Xie}}, \bibinfo {author} {\bibfnamefont {T.}~\bibnamefont
  {Aoki}}, \bibinfo {author} {\bibfnamefont {L.}~\bibnamefont {Li}}, \bibinfo
  {author} {\bibfnamefont {C.}~\bibnamefont {Heikes}}, \bibinfo {author}
  {\bibfnamefont {S.~P.}\ \bibnamefont {Lau}}, \bibinfo {author} {\bibfnamefont
  {D.~G.}\ \bibnamefont {Schlom}}, \emph {et~al.},\ }\bibfield  {title}
  {\bibinfo {title} {Intrinsic conductance of domain walls in {BiFeO$_3$}},\
  }\href@noop {} {\bibfield  {journal} {\bibinfo  {journal} {Advanced
  Materials}\ }\textbf {\bibinfo {volume} {31}},\ \bibinfo {pages} {1902099}
  (\bibinfo {year} {2019}{\natexlab{a}})}\BibitemShut {NoStop}%
\bibitem [{\citenamefont {Sluka}\ \emph {et~al.}(2013)\citenamefont {Sluka},
  \citenamefont {Tagantsev}, \citenamefont {Bednyakov},\ and\ \citenamefont
  {Setter}}]{sluka2013free}%
  \BibitemOpen
  \bibfield  {author} {\bibinfo {author} {\bibfnamefont {T.}~\bibnamefont
  {Sluka}}, \bibinfo {author} {\bibfnamefont {A.~K.}\ \bibnamefont
  {Tagantsev}}, \bibinfo {author} {\bibfnamefont {P.}~\bibnamefont
  {Bednyakov}},\ and\ \bibinfo {author} {\bibfnamefont {N.}~\bibnamefont
  {Setter}},\ }\bibfield  {title} {\bibinfo {title} {Free-electron gas at
  charged domain walls in insulating {BaTiO$_3$}},\ }\href
  {https://doi.org/10.1038/ncomms2813} {\bibfield  {journal} {\bibinfo
  {journal} {Nature Communications}\ }\textbf {\bibinfo {volume} {4}},\
  \bibinfo {pages} {1808} (\bibinfo {year} {2013})}\BibitemShut {NoStop}%
\bibitem [{\citenamefont {Beccard}\ \emph {et~al.}(2022)\citenamefont
  {Beccard}, \citenamefont {Kirbus}, \citenamefont {Beyreuther}, \citenamefont
  {Rüsing}, \citenamefont {Bednyakov}, \citenamefont {Hlinka},\ and\
  \citenamefont {Eng}}]{Beccard:2022}%
  \BibitemOpen
  \bibfield  {author} {\bibinfo {author} {\bibfnamefont {H.}~\bibnamefont
  {Beccard}}, \bibinfo {author} {\bibfnamefont {B.}~\bibnamefont {Kirbus}},
  \bibinfo {author} {\bibfnamefont {E.}~\bibnamefont {Beyreuther}}, \bibinfo
  {author} {\bibfnamefont {M.}~\bibnamefont {Rüsing}}, \bibinfo {author}
  {\bibfnamefont {P.}~\bibnamefont {Bednyakov}}, \bibinfo {author}
  {\bibfnamefont {J.}~\bibnamefont {Hlinka}},\ and\ \bibinfo {author}
  {\bibfnamefont {L.~M.}\ \bibnamefont {Eng}},\ }\bibfield  {title} {\bibinfo
  {title} {Nanoscale conductive sheets in ferroelectric {BaTiO$_3$}: Large hall
  electron mobilities at head-to-head domain walls},\ }\href@noop {} {\bibfield
   {journal} {\bibinfo  {journal} {ACS Applied Nano Materials}\ }\textbf
  {\bibinfo {volume} {5}},\ \bibinfo {pages} {8717} (\bibinfo {year}
  {2022})}\BibitemShut {NoStop}%
\bibitem [{\citenamefont {Schr{\"o}der}\ \emph {et~al.}(2012)\citenamefont
  {Schr{\"o}der}, \citenamefont {Hau{\ss}mann}, \citenamefont {Thiessen},
  \citenamefont {Soergel}, \citenamefont {Woike},\ and\ \citenamefont
  {Eng}}]{schroder2012conducting}%
  \BibitemOpen
  \bibfield  {author} {\bibinfo {author} {\bibfnamefont {M.}~\bibnamefont
  {Schr{\"o}der}}, \bibinfo {author} {\bibfnamefont {A.}~\bibnamefont
  {Hau{\ss}mann}}, \bibinfo {author} {\bibfnamefont {A.}~\bibnamefont
  {Thiessen}}, \bibinfo {author} {\bibfnamefont {E.}~\bibnamefont {Soergel}},
  \bibinfo {author} {\bibfnamefont {T.}~\bibnamefont {Woike}},\ and\ \bibinfo
  {author} {\bibfnamefont {L.~M.}\ \bibnamefont {Eng}},\ }\bibfield  {title}
  {\bibinfo {title} {Conducting domain walls in lithium niobate single
  crystals},\ }\href@noop {} {\bibfield  {journal} {\bibinfo  {journal}
  {Advanced functional materials}\ }\textbf {\bibinfo {volume} {22}},\ \bibinfo
  {pages} {3936} (\bibinfo {year} {2012})}\BibitemShut {NoStop}%
\bibitem [{\citenamefont {Werner}\ \emph {et~al.}(2017)\citenamefont {Werner},
  \citenamefont {Herr}, \citenamefont {Buse}, \citenamefont {Sturman},
  \citenamefont {Soergel}, \citenamefont {Razzaghi},\ and\ \citenamefont
  {Breunig}}]{werner2017large}%
  \BibitemOpen
  \bibfield  {author} {\bibinfo {author} {\bibfnamefont {C.~S.}\ \bibnamefont
  {Werner}}, \bibinfo {author} {\bibfnamefont {S.~J.}\ \bibnamefont {Herr}},
  \bibinfo {author} {\bibfnamefont {K.}~\bibnamefont {Buse}}, \bibinfo {author}
  {\bibfnamefont {B.}~\bibnamefont {Sturman}}, \bibinfo {author} {\bibfnamefont
  {E.}~\bibnamefont {Soergel}}, \bibinfo {author} {\bibfnamefont
  {C.}~\bibnamefont {Razzaghi}},\ and\ \bibinfo {author} {\bibfnamefont
  {I.}~\bibnamefont {Breunig}},\ }\bibfield  {title} {\bibinfo {title} {Large
  and accessible conductivity of charged domain walls in lithium niobate},\
  }\href@noop {} {\bibfield  {journal} {\bibinfo  {journal} {Scientific
  Reports}\ }\textbf {\bibinfo {volume} {7}},\ \bibinfo {pages} {9862}
  (\bibinfo {year} {2017})}\BibitemShut {NoStop}%
\bibitem [{\citenamefont {Volk}\ \emph {et~al.}(2017)\citenamefont {Volk},
  \citenamefont {Gainutdinov},\ and\ \citenamefont {Zhang}}]{volk2017domain}%
  \BibitemOpen
  \bibfield  {author} {\bibinfo {author} {\bibfnamefont {T.}~\bibnamefont
  {Volk}}, \bibinfo {author} {\bibfnamefont {R.}~\bibnamefont {Gainutdinov}},\
  and\ \bibinfo {author} {\bibfnamefont {H.}~\bibnamefont {Zhang}},\ }\bibfield
   {title} {\bibinfo {title} {Domain-wall conduction in {AFM}-written domain
  patterns in ion-sliced {LiNbO$_3$} films},\ }\href@noop {} {\bibfield
  {journal} {\bibinfo  {journal} {Applied Physics Letters}\ }\textbf {\bibinfo
  {volume} {110}} (\bibinfo {year} {2017})}\BibitemShut {NoStop}%
\bibitem [{\citenamefont {Qian}\ \emph {et~al.}(2022)\citenamefont {Qian},
  \citenamefont {Zhang}, \citenamefont {Xu},\ and\ \citenamefont
  {Zhang}}]{Qian:2022}%
  \BibitemOpen
  \bibfield  {author} {\bibinfo {author} {\bibfnamefont {Y.}~\bibnamefont
  {Qian}}, \bibinfo {author} {\bibfnamefont {Y.}~\bibnamefont {Zhang}},
  \bibinfo {author} {\bibfnamefont {J.}~\bibnamefont {Xu}},\ and\ \bibinfo
  {author} {\bibfnamefont {G.}~\bibnamefont {Zhang}},\ }\bibfield  {title}
  {\bibinfo {title} {Domain-wall $p$-$n$ junction in lithium niobate thin film
  on an insulator},\ }\href {https://doi.org/10.1103/PhysRevApplied.17.044011}
  {\bibfield  {journal} {\bibinfo  {journal} {Phys. Rev. Appl.}\ }\textbf
  {\bibinfo {volume} {17}},\ \bibinfo {pages} {044011} (\bibinfo {year}
  {2022})}\BibitemShut {NoStop}%
\bibitem [{\citenamefont {McCluskey}\ \emph
  {et~al.}(2025{\natexlab{a}})\citenamefont {McCluskey}, \citenamefont
  {Dalzell}, \citenamefont {Kumar},\ and\ \citenamefont
  {Gregg}}]{McClusky:2025}%
  \BibitemOpen
  \bibfield  {author} {\bibinfo {author} {\bibfnamefont {C.~J.}\ \bibnamefont
  {McCluskey}}, \bibinfo {author} {\bibfnamefont {J.}~\bibnamefont {Dalzell}},
  \bibinfo {author} {\bibfnamefont {A.}~\bibnamefont {Kumar}},\ and\ \bibinfo
  {author} {\bibfnamefont {J.~M.}\ \bibnamefont {Gregg}},\ }\bibfield  {title}
  {\bibinfo {title} {Current flow mapping in conducting ferroelectric domain
  walls using scanning {NV}-magnetometry},\ }\href@noop {} {\bibfield
  {journal} {\bibinfo  {journal} {Advanced Electronic Materials}\ ,\ \bibinfo
  {pages} {e00142}} (\bibinfo {year} {2025}{\natexlab{a}})},\ \Eprint
  {https://arxiv.org/abs/https://doi.org/10.1002/aelm.202500142}
  {https://doi.org/10.1002/aelm.202500142} \BibitemShut {NoStop}%
\bibitem [{\citenamefont {Sharma}\ \emph {et~al.}(2017)\citenamefont {Sharma},
  \citenamefont {Zhang}, \citenamefont {Sando}, \citenamefont {Lei},
  \citenamefont {Liu}, \citenamefont {Li}, \citenamefont {Nagarajan},\ and\
  \citenamefont {Seidel}}]{Sharma:2017}%
  \BibitemOpen
  \bibfield  {author} {\bibinfo {author} {\bibfnamefont {P.}~\bibnamefont
  {Sharma}}, \bibinfo {author} {\bibfnamefont {Q.}~\bibnamefont {Zhang}},
  \bibinfo {author} {\bibfnamefont {D.}~\bibnamefont {Sando}}, \bibinfo
  {author} {\bibfnamefont {C.~H.}\ \bibnamefont {Lei}}, \bibinfo {author}
  {\bibfnamefont {Y.}~\bibnamefont {Liu}}, \bibinfo {author} {\bibfnamefont
  {J.}~\bibnamefont {Li}}, \bibinfo {author} {\bibfnamefont {V.}~\bibnamefont
  {Nagarajan}},\ and\ \bibinfo {author} {\bibfnamefont {J.}~\bibnamefont
  {Seidel}},\ }\bibfield  {title} {\bibinfo {title} {Nonvolatile ferroelectric
  domain wall memory},\ }\href {https://doi.org/10.1126/sciadv.1700512}
  {\bibfield  {journal} {\bibinfo  {journal} {Science Advances}\ }\textbf
  {\bibinfo {volume} {3}},\ \bibinfo {pages} {e1700512} (\bibinfo {year}
  {2017})}\BibitemShut {NoStop}%
\bibitem [{\citenamefont {Ma}\ \emph {et~al.}(2018)\citenamefont {Ma},
  \citenamefont {Ma}, \citenamefont {Zhang}, \citenamefont {Peng},
  \citenamefont {Wang}, \citenamefont {Liu}, \citenamefont {Wang},
  \citenamefont {Li}, \citenamefont {Chen}, \citenamefont {Cheng} \emph
  {et~al.}}]{Ma:2018}%
  \BibitemOpen
  \bibfield  {author} {\bibinfo {author} {\bibfnamefont {J.}~\bibnamefont
  {Ma}}, \bibinfo {author} {\bibfnamefont {J.}~\bibnamefont {Ma}}, \bibinfo
  {author} {\bibfnamefont {Q.}~\bibnamefont {Zhang}}, \bibinfo {author}
  {\bibfnamefont {R.}~\bibnamefont {Peng}}, \bibinfo {author} {\bibfnamefont
  {J.}~\bibnamefont {Wang}}, \bibinfo {author} {\bibfnamefont {C.}~\bibnamefont
  {Liu}}, \bibinfo {author} {\bibfnamefont {M.}~\bibnamefont {Wang}}, \bibinfo
  {author} {\bibfnamefont {N.}~\bibnamefont {Li}}, \bibinfo {author}
  {\bibfnamefont {M.}~\bibnamefont {Chen}}, \bibinfo {author} {\bibfnamefont
  {X.}~\bibnamefont {Cheng}}, \emph {et~al.},\ }\bibfield  {title} {\bibinfo
  {title} {Controllable conductive readout in self-assembled, topologically
  confined ferroelectric domain walls},\ }\href@noop {} {\bibfield  {journal}
  {\bibinfo  {journal} {Nature nanotechnology}\ }\textbf {\bibinfo {volume}
  {13}},\ \bibinfo {pages} {947} (\bibinfo {year} {2018})}\BibitemShut
  {NoStop}%
\bibitem [{\citenamefont {Jiang}\ \emph {et~al.}(2018)\citenamefont {Jiang},
  \citenamefont {Bai}, \citenamefont {Chen}, \citenamefont {He}, \citenamefont
  {Zhang}, \citenamefont {Zhang}, \citenamefont {Shi}, \citenamefont {Park},
  \citenamefont {Scott}, \citenamefont {Hwang} \emph {et~al.}}]{Jiang:2018}%
  \BibitemOpen
  \bibfield  {author} {\bibinfo {author} {\bibfnamefont {J.}~\bibnamefont
  {Jiang}}, \bibinfo {author} {\bibfnamefont {Z.~L.}\ \bibnamefont {Bai}},
  \bibinfo {author} {\bibfnamefont {Z.~H.}\ \bibnamefont {Chen}}, \bibinfo
  {author} {\bibfnamefont {L.}~\bibnamefont {He}}, \bibinfo {author}
  {\bibfnamefont {D.~W.}\ \bibnamefont {Zhang}}, \bibinfo {author}
  {\bibfnamefont {Q.~H.}\ \bibnamefont {Zhang}}, \bibinfo {author}
  {\bibfnamefont {J.~A.}\ \bibnamefont {Shi}}, \bibinfo {author} {\bibfnamefont
  {M.~H.}\ \bibnamefont {Park}}, \bibinfo {author} {\bibfnamefont {J.~F.}\
  \bibnamefont {Scott}}, \bibinfo {author} {\bibfnamefont {C.~S.}\ \bibnamefont
  {Hwang}}, \emph {et~al.},\ }\bibfield  {title} {\bibinfo {title} {Temporary
  formation of highly conducting domain walls for non-destructive read-out of
  ferroelectric domain-wall resistance switching memories},\ }\href@noop {}
  {\bibfield  {journal} {\bibinfo  {journal} {Nature materials}\ }\textbf
  {\bibinfo {volume} {17}},\ \bibinfo {pages} {49} (\bibinfo {year}
  {2018})}\BibitemShut {NoStop}%
\bibitem [{\citenamefont {Sharma}\ \emph {et~al.}(2019)\citenamefont {Sharma},
  \citenamefont {Sando}, \citenamefont {Zhang}, \citenamefont {Cheng},
  \citenamefont {Prosandeev}, \citenamefont {Bulanadi}, \citenamefont
  {Prokhorenko}, \citenamefont {Bellaiche}, \citenamefont {Chen}, \citenamefont
  {Nagarajan},\ and\ \citenamefont {Seidel}}]{Sharma:2019}%
  \BibitemOpen
  \bibfield  {author} {\bibinfo {author} {\bibfnamefont {P.}~\bibnamefont
  {Sharma}}, \bibinfo {author} {\bibfnamefont {D.}~\bibnamefont {Sando}},
  \bibinfo {author} {\bibfnamefont {Q.}~\bibnamefont {Zhang}}, \bibinfo
  {author} {\bibfnamefont {X.}~\bibnamefont {Cheng}}, \bibinfo {author}
  {\bibfnamefont {S.}~\bibnamefont {Prosandeev}}, \bibinfo {author}
  {\bibfnamefont {R.}~\bibnamefont {Bulanadi}}, \bibinfo {author}
  {\bibfnamefont {S.}~\bibnamefont {Prokhorenko}}, \bibinfo {author}
  {\bibfnamefont {L.}~\bibnamefont {Bellaiche}}, \bibinfo {author}
  {\bibfnamefont {L.-Q.}\ \bibnamefont {Chen}}, \bibinfo {author}
  {\bibfnamefont {V.}~\bibnamefont {Nagarajan}},\ and\ \bibinfo {author}
  {\bibfnamefont {J.}~\bibnamefont {Seidel}},\ }\bibfield  {title} {\bibinfo
  {title} {Conformational domain wall switch},\ }\href
  {https://doi.org/https://doi.org/10.1002/adfm.201807523} {\bibfield
  {journal} {\bibinfo  {journal} {Advanced Functional Materials}\ }\textbf
  {\bibinfo {volume} {29}},\ \bibinfo {pages} {1807523} (\bibinfo {year}
  {2019})}\BibitemShut {NoStop}%
\bibitem [{\citenamefont {McConville}\ \emph {et~al.}(2020)\citenamefont
  {McConville}, \citenamefont {Lu}, \citenamefont {Wang}, \citenamefont {Tan},
  \citenamefont {Cochard}, \citenamefont {Conroy}, \citenamefont {Moore},
  \citenamefont {Harvey}, \citenamefont {Bangert}, \citenamefont {Chen},
  \citenamefont {Gruverman},\ and\ \citenamefont {Gregg}}]{McConville:2020}%
  \BibitemOpen
  \bibfield  {author} {\bibinfo {author} {\bibfnamefont {J.~P.~V.}\
  \bibnamefont {McConville}}, \bibinfo {author} {\bibfnamefont
  {H.}~\bibnamefont {Lu}}, \bibinfo {author} {\bibfnamefont {B.}~\bibnamefont
  {Wang}}, \bibinfo {author} {\bibfnamefont {Y.}~\bibnamefont {Tan}}, \bibinfo
  {author} {\bibfnamefont {C.}~\bibnamefont {Cochard}}, \bibinfo {author}
  {\bibfnamefont {M.}~\bibnamefont {Conroy}}, \bibinfo {author} {\bibfnamefont
  {K.}~\bibnamefont {Moore}}, \bibinfo {author} {\bibfnamefont
  {A.}~\bibnamefont {Harvey}}, \bibinfo {author} {\bibfnamefont
  {U.}~\bibnamefont {Bangert}}, \bibinfo {author} {\bibfnamefont {L.-Q.}\
  \bibnamefont {Chen}}, \bibinfo {author} {\bibfnamefont {A.}~\bibnamefont
  {Gruverman}},\ and\ \bibinfo {author} {\bibfnamefont {J.~M.}\ \bibnamefont
  {Gregg}},\ }\bibfield  {title} {\bibinfo {title} {Ferroelectric domain wall
  memristor},\ }\href {https://doi.org/https://doi.org/10.1002/adfm.202000109}
  {\bibfield  {journal} {\bibinfo  {journal} {Advanced Functional Materials}\
  }\textbf {\bibinfo {volume} {30}},\ \bibinfo {pages} {2000109} (\bibinfo
  {year} {2020})}\BibitemShut {NoStop}%
\bibitem [{\citenamefont {Risch}\ \emph {et~al.}(2022)\citenamefont {Risch},
  \citenamefont {Tikhonov}, \citenamefont {Lukyanchuk}, \citenamefont
  {Ionescu},\ and\ \citenamefont {Stolichnov}}]{Risch:2022}%
  \BibitemOpen
  \bibfield  {author} {\bibinfo {author} {\bibfnamefont {F.}~\bibnamefont
  {Risch}}, \bibinfo {author} {\bibfnamefont {Y.}~\bibnamefont {Tikhonov}},
  \bibinfo {author} {\bibfnamefont {I.}~\bibnamefont {Lukyanchuk}}, \bibinfo
  {author} {\bibfnamefont {A.~M.}\ \bibnamefont {Ionescu}},\ and\ \bibinfo
  {author} {\bibfnamefont {I.}~\bibnamefont {Stolichnov}},\ }\bibfield  {title}
  {\bibinfo {title} {Giant switchable non thermally-activated conduction in
  180° domain walls in tetragonal {Pb(Zr,Ti)O$_3$}},\ }\href@noop {}
  {\bibfield  {journal} {\bibinfo  {journal} {Nature Communications}\ }\textbf
  {\bibinfo {volume} {13}},\ \bibinfo {pages} {7239} (\bibinfo {year}
  {2022})}\BibitemShut {NoStop}%
\bibitem [{\citenamefont {Wang}\ \emph {et~al.}(2022)\citenamefont {Wang},
  \citenamefont {Ma}, \citenamefont {Huang}, \citenamefont {Ma}, \citenamefont
  {Jafri}, \citenamefont {Fan}, \citenamefont {Yang}, \citenamefont {Wang},
  \citenamefont {Chen}, \citenamefont {Liu} \emph {et~al.}}]{Wang:2022}%
  \BibitemOpen
  \bibfield  {author} {\bibinfo {author} {\bibfnamefont {J.}~\bibnamefont
  {Wang}}, \bibinfo {author} {\bibfnamefont {J.}~\bibnamefont {Ma}}, \bibinfo
  {author} {\bibfnamefont {H.}~\bibnamefont {Huang}}, \bibinfo {author}
  {\bibfnamefont {J.}~\bibnamefont {Ma}}, \bibinfo {author} {\bibfnamefont
  {H.~M.}\ \bibnamefont {Jafri}}, \bibinfo {author} {\bibfnamefont
  {Y.}~\bibnamefont {Fan}}, \bibinfo {author} {\bibfnamefont {H.}~\bibnamefont
  {Yang}}, \bibinfo {author} {\bibfnamefont {Y.}~\bibnamefont {Wang}}, \bibinfo
  {author} {\bibfnamefont {M.}~\bibnamefont {Chen}}, \bibinfo {author}
  {\bibfnamefont {D.}~\bibnamefont {Liu}}, \emph {et~al.},\ }\bibfield  {title}
  {\bibinfo {title} {Ferroelectric domain-wall logic units},\ }\href@noop {}
  {\bibfield  {journal} {\bibinfo  {journal} {Nature communications}\ }\textbf
  {\bibinfo {volume} {13}},\ \bibinfo {pages} {3255} (\bibinfo {year}
  {2022})}\BibitemShut {NoStop}%
\bibitem [{\citenamefont {{Liu}}\ \emph {et~al.}(2023)\citenamefont {{Liu}},
  \citenamefont {{Wang}}, \citenamefont {{Li}}, \citenamefont {{Tao}},
  \citenamefont {{Paudel}}, \citenamefont {{Yu}}, \citenamefont {{Wang}},
  \citenamefont {{Hong}}, \citenamefont {{Zhang}}, \citenamefont {{Ren}},
  \citenamefont {{Xie}}, \citenamefont {{Tsymbal}}, \citenamefont {{Chen}},
  \citenamefont {{Zhang}},\ and\ \citenamefont {{Tian}}}]{Liu:2023}%
  \BibitemOpen
  \bibfield  {author} {\bibinfo {author} {\bibfnamefont {Z.}~\bibnamefont
  {{Liu}}}, \bibinfo {author} {\bibfnamefont {H.}~\bibnamefont {{Wang}}},
  \bibinfo {author} {\bibfnamefont {M.}~\bibnamefont {{Li}}}, \bibinfo {author}
  {\bibfnamefont {L.}~\bibnamefont {{Tao}}}, \bibinfo {author} {\bibfnamefont
  {T.~R.}\ \bibnamefont {{Paudel}}}, \bibinfo {author} {\bibfnamefont
  {H.}~\bibnamefont {{Yu}}}, \bibinfo {author} {\bibfnamefont {Y.}~\bibnamefont
  {{Wang}}}, \bibinfo {author} {\bibfnamefont {S.}~\bibnamefont {{Hong}}},
  \bibinfo {author} {\bibfnamefont {M.}~\bibnamefont {{Zhang}}}, \bibinfo
  {author} {\bibfnamefont {Z.}~\bibnamefont {{Ren}}}, \bibinfo {author}
  {\bibfnamefont {Y.}~\bibnamefont {{Xie}}}, \bibinfo {author} {\bibfnamefont
  {E.~Y.}\ \bibnamefont {{Tsymbal}}}, \bibinfo {author} {\bibfnamefont
  {J.}~\bibnamefont {{Chen}}}, \bibinfo {author} {\bibfnamefont
  {Z.}~\bibnamefont {{Zhang}}},\ and\ \bibinfo {author} {\bibfnamefont
  {H.}~\bibnamefont {{Tian}}},\ }\bibfield  {title} {\bibinfo {title}
  {{In-plane charged domain walls with memristive behaviour in a ferroelectric
  film}},\ }\href {https://doi.org/10.1038/s41586-022-05503-5} {\bibfield
  {journal} {\bibinfo  {journal} {\nat}\ }\textbf {\bibinfo {volume} {613}},\
  \bibinfo {pages} {656} (\bibinfo {year} {2023})}\BibitemShut {NoStop}%
\bibitem [{\citenamefont {Ratzenberger}\ \emph {et~al.}(2024)\citenamefont
  {Ratzenberger}, \citenamefont {Kiseleva}, \citenamefont {Koppitz},
  \citenamefont {Beyreuther}, \citenamefont {Zahn}, \citenamefont {Gössel},
  \citenamefont {Hegarty}, \citenamefont {Amber}, \citenamefont {Rüsing},\
  and\ \citenamefont {Eng}}]{Ratzenberger:2024}%
  \BibitemOpen
  \bibfield  {author} {\bibinfo {author} {\bibfnamefont {J.}~\bibnamefont
  {Ratzenberger}}, \bibinfo {author} {\bibfnamefont {I.}~\bibnamefont
  {Kiseleva}}, \bibinfo {author} {\bibfnamefont {B.}~\bibnamefont {Koppitz}},
  \bibinfo {author} {\bibfnamefont {E.}~\bibnamefont {Beyreuther}}, \bibinfo
  {author} {\bibfnamefont {M.}~\bibnamefont {Zahn}}, \bibinfo {author}
  {\bibfnamefont {J.}~\bibnamefont {Gössel}}, \bibinfo {author} {\bibfnamefont
  {P.~A.}\ \bibnamefont {Hegarty}}, \bibinfo {author} {\bibfnamefont {Z.~H.}\
  \bibnamefont {Amber}}, \bibinfo {author} {\bibfnamefont {M.}~\bibnamefont
  {Rüsing}},\ and\ \bibinfo {author} {\bibfnamefont {L.~M.}\ \bibnamefont
  {Eng}},\ }\bibfield  {title} {\bibinfo {title} {Toward the reproducible
  fabrication of conductive ferroelectric domain walls into lithium niobate
  bulk single crystals},\ }\href {https://doi.org/10.1063/5.0219300} {\bibfield
   {journal} {\bibinfo  {journal} {Journal of Applied Physics}\ }\textbf
  {\bibinfo {volume} {136}},\ \bibinfo {pages} {104302} (\bibinfo {year}
  {2024})}\BibitemShut {NoStop}%
\bibitem [{\citenamefont {Sharma}\ \emph {et~al.}(2022)\citenamefont {Sharma},
  \citenamefont {Moise}, \citenamefont {Colombo},\ and\ \citenamefont
  {Seidel}}]{sharma2022roadmap}%
  \BibitemOpen
  \bibfield  {author} {\bibinfo {author} {\bibfnamefont {P.}~\bibnamefont
  {Sharma}}, \bibinfo {author} {\bibfnamefont {T.~S.}\ \bibnamefont {Moise}},
  \bibinfo {author} {\bibfnamefont {L.}~\bibnamefont {Colombo}},\ and\ \bibinfo
  {author} {\bibfnamefont {J.}~\bibnamefont {Seidel}},\ }\bibfield  {title}
  {\bibinfo {title} {Roadmap for ferroelectric domain wall nanoelectronics},\
  }\href@noop {} {\bibfield  {journal} {\bibinfo  {journal} {Advanced
  Functional Materials}\ }\textbf {\bibinfo {volume} {32}},\ \bibinfo {pages}
  {2110263} (\bibinfo {year} {2022})}\BibitemShut {NoStop}%
\bibitem [{\citenamefont {Sturman}\ and\ \citenamefont
  {Podivilov}(2023)}]{sturman2023effect}%
  \BibitemOpen
  \bibfield  {author} {\bibinfo {author} {\bibfnamefont {B.}~\bibnamefont
  {Sturman}}\ and\ \bibinfo {author} {\bibfnamefont {E.}~\bibnamefont
  {Podivilov}},\ }\bibfield  {title} {\bibinfo {title} {Effect of domain wall
  conductivity on domain formation energy},\ }\href@noop {} {\bibfield
  {journal} {\bibinfo  {journal} {Ferroelectrics}\ }\textbf {\bibinfo {volume}
  {604}},\ \bibinfo {pages} {80} (\bibinfo {year} {2023})}\BibitemShut
  {NoStop}%
\bibitem [{\citenamefont {Kosobokov}\ \emph {et~al.}(2024)\citenamefont
  {Kosobokov}, \citenamefont {Turygin}, \citenamefont {Melnikov}, \citenamefont
  {Shur},\ and\ \citenamefont {Alikin}}]{Kosobokov:2024conductivity}%
  \BibitemOpen
  \bibfield  {author} {\bibinfo {author} {\bibfnamefont {M.}~\bibnamefont
  {Kosobokov}}, \bibinfo {author} {\bibfnamefont {A.}~\bibnamefont {Turygin}},
  \bibinfo {author} {\bibfnamefont {S.}~\bibnamefont {Melnikov}}, \bibinfo
  {author} {\bibfnamefont {V.}~\bibnamefont {Shur}},\ and\ \bibinfo {author}
  {\bibfnamefont {D.}~\bibnamefont {Alikin}},\ }\bibfield  {title} {\bibinfo
  {title} {Role of domain wall conductivity in the stability of ferroelectric
  domains in ferroelectric single crystals},\ }\href
  {https://doi.org/10.1103/PhysRevB.110.134116} {\bibfield  {journal} {\bibinfo
   {journal} {Phys. Rev. B}\ }\textbf {\bibinfo {volume} {110}},\ \bibinfo
  {pages} {134116} (\bibinfo {year} {2024})}\BibitemShut {NoStop}%
\bibitem [{\citenamefont {Bednyakov}\ \emph {et~al.}(2025)\citenamefont
  {Bednyakov}, \citenamefont {Rafalovskyi},\ and\ \citenamefont
  {Hlinka}}]{bednyakov2025fragmented}%
  \BibitemOpen
  \bibfield  {author} {\bibinfo {author} {\bibfnamefont {P.~S.}\ \bibnamefont
  {Bednyakov}}, \bibinfo {author} {\bibfnamefont {I.}~\bibnamefont
  {Rafalovskyi}},\ and\ \bibinfo {author} {\bibfnamefont {J.}~\bibnamefont
  {Hlinka}},\ }\bibfield  {title} {\bibinfo {title} {Fragmented charged domain
  wall below the tetragonal-orthorhombic phase transition in {BaTiO$_3$}},\
  }\href@noop {} {\bibfield  {journal} {\bibinfo  {journal} {Applied Physics
  Letters}\ }\textbf {\bibinfo {volume} {126}} (\bibinfo {year}
  {2025})}\BibitemShut {NoStop}%
\bibitem [{\citenamefont {Yang}(2024)}]{yang2024intrinsic}%
  \BibitemOpen
  \bibfield  {author} {\bibinfo {author} {\bibfnamefont {F.}~\bibnamefont
  {Yang}},\ }\bibfield  {title} {\bibinfo {title} {Intrinsic conductance of
  ferroelectric charged domain walls},\ }\href@noop {} {\bibfield  {journal}
  {\bibinfo  {journal} {Physics}\ }\textbf {\bibinfo {volume} {6}},\ \bibinfo
  {pages} {1083} (\bibinfo {year} {2024})}\BibitemShut {NoStop}%
\bibitem [{\citenamefont {Marton}\ \emph {et~al.}(2025)\citenamefont {Marton},
  \citenamefont {Paściak}, \citenamefont {Gonçalves}, \citenamefont {Novák},
  \citenamefont {Hlinka}, \citenamefont {Beanland},\ and\ \citenamefont
  {Alexe}}]{marton2025pyramidal}%
  \BibitemOpen
  \bibfield  {author} {\bibinfo {author} {\bibfnamefont {P.}~\bibnamefont
  {Marton}}, \bibinfo {author} {\bibfnamefont {M.}~\bibnamefont {Paściak}},
  \bibinfo {author} {\bibfnamefont {M.}~\bibnamefont {Gonçalves}}, \bibinfo
  {author} {\bibfnamefont {O.}~\bibnamefont {Novák}}, \bibinfo {author}
  {\bibfnamefont {J.}~\bibnamefont {Hlinka}}, \bibinfo {author} {\bibfnamefont
  {R.}~\bibnamefont {Beanland}},\ and\ \bibinfo {author} {\bibfnamefont
  {M.}~\bibnamefont {Alexe}},\ }\href {https://arxiv.org/abs/2501.01190}
  {\bibinfo {title} {Pyramidal charged domain walls in ferroelectric
  {BiFeO$_3$}}} (\bibinfo {year} {2025}),\ \Eprint
  {https://arxiv.org/abs/2501.01190} {arXiv:2501.01190 [cond-mat.mtrl-sci]}
  \BibitemShut {NoStop}%
\bibitem [{\citenamefont {Khachaturyan}\ \emph {et~al.}(2024)\citenamefont
  {Khachaturyan}, \citenamefont {Yang}, \citenamefont {Teng}, \citenamefont
  {Udofia}, \citenamefont {Stricker},\ and\ \citenamefont
  {Gr\"unebohm}}]{Khachaturyan:2024}%
  \BibitemOpen
  \bibfield  {author} {\bibinfo {author} {\bibfnamefont {R.}~\bibnamefont
  {Khachaturyan}}, \bibinfo {author} {\bibfnamefont {Y.}~\bibnamefont {Yang}},
  \bibinfo {author} {\bibfnamefont {S.-H.}\ \bibnamefont {Teng}}, \bibinfo
  {author} {\bibfnamefont {B.}~\bibnamefont {Udofia}}, \bibinfo {author}
  {\bibfnamefont {M.}~\bibnamefont {Stricker}},\ and\ \bibinfo {author}
  {\bibfnamefont {A.}~\bibnamefont {Gr\"unebohm}},\ }\bibfield  {title}
  {\bibinfo {title} {Microscopic insights on field induced switching and domain
  wall motion in orthorhombic ferroelectrics},\ }\href
  {https://doi.org/10.1103/PhysRevMaterials.8.024403} {\bibfield  {journal}
  {\bibinfo  {journal} {Phys. Rev. Mater.}\ }\textbf {\bibinfo {volume} {8}},\
  \bibinfo {pages} {024403} (\bibinfo {year} {2024})}\BibitemShut {NoStop}%
\bibitem [{\citenamefont {Carroll}\ and\ \citenamefont
  {Atkinson}(2025)}]{carroll2025dynamics}%
  \BibitemOpen
  \bibfield  {author} {\bibinfo {author} {\bibfnamefont {C.}~\bibnamefont
  {Carroll}}\ and\ \bibinfo {author} {\bibfnamefont {W.~A.}\ \bibnamefont
  {Atkinson}},\ }\bibfield  {title} {\bibinfo {title} {Dynamics of conducting
  ferroelectric domain walls},\ }\href {https://doi.org/10.1103/v293-1x2y}
  {\bibfield  {journal} {\bibinfo  {journal} {Phys. Rev. B}\ }\textbf {\bibinfo
  {volume} {112}},\ \bibinfo {pages} {024316} (\bibinfo {year}
  {2025})}\BibitemShut {NoStop}%
\bibitem [{\citenamefont {Gurung}\ \emph {et~al.}(2025)\citenamefont {Gurung},
  \citenamefont {Ishtiyaq}, \citenamefont {Alpay}, \citenamefont {Mangeri},\
  and\ \citenamefont {Nakhmanson}}]{Gurung:2025extrinsic}%
  \BibitemOpen
  \bibfield  {author} {\bibinfo {author} {\bibfnamefont {A.}~\bibnamefont
  {Gurung}}, \bibinfo {author} {\bibfnamefont {M.~F.}\ \bibnamefont
  {Ishtiyaq}}, \bibinfo {author} {\bibfnamefont {S.~P.}\ \bibnamefont {Alpay}},
  \bibinfo {author} {\bibfnamefont {J.}~\bibnamefont {Mangeri}},\ and\ \bibinfo
  {author} {\bibfnamefont {S.}~\bibnamefont {Nakhmanson}},\ }\bibfield  {title}
  {\bibinfo {title} {Extrinsic dielectric response due to domain wall motion in
  ferroelectric {BaTiO$_3$}},\ }\href@noop {} {\bibfield  {journal} {\bibinfo
  {journal} {Computational Materials Today}\ }\textbf {\bibinfo {volume} {5}},\
  \bibinfo {pages} {100016} (\bibinfo {year} {2025})}\BibitemShut {NoStop}%
\bibitem [{\citenamefont {McCluskey}\ \emph
  {et~al.}(2025{\natexlab{b}})\citenamefont {McCluskey}, \citenamefont
  {Holsgrove},\ and\ \citenamefont {Gregg}}]{McCluskey:2025Review}%
  \BibitemOpen
  \bibfield  {author} {\bibinfo {author} {\bibfnamefont {C.~J.}\ \bibnamefont
  {McCluskey}}, \bibinfo {author} {\bibfnamefont {K.~M.}\ \bibnamefont
  {Holsgrove}},\ and\ \bibinfo {author} {\bibfnamefont {J.~M.}\ \bibnamefont
  {Gregg}},\ }\bibfield  {title} {\bibinfo {title} {Perspective: Domain wall
  nanoelectronics},\ }\href {https://doi.org/10.1063/5.0279244} {\bibfield
  {journal} {\bibinfo  {journal} {Journal of Applied Physics}\ }\textbf
  {\bibinfo {volume} {138}},\ \bibinfo {pages} {050901} (\bibinfo {year}
  {2025}{\natexlab{b}})}\BibitemShut {NoStop}%
\bibitem [{\citenamefont {Sifuna}\ \emph {et~al.}(2020)\citenamefont {Sifuna},
  \citenamefont {Garc\'{\i}a-Fern\'andez}, \citenamefont {Manyali},
  \citenamefont {Amolo},\ and\ \citenamefont {Junquera}}]{sifuna:2020DFT}%
  \BibitemOpen
  \bibfield  {author} {\bibinfo {author} {\bibfnamefont {J.}~\bibnamefont
  {Sifuna}}, \bibinfo {author} {\bibfnamefont {P.}~\bibnamefont
  {Garc\'{\i}a-Fern\'andez}}, \bibinfo {author} {\bibfnamefont {G.~S.}\
  \bibnamefont {Manyali}}, \bibinfo {author} {\bibfnamefont {G.}~\bibnamefont
  {Amolo}},\ and\ \bibinfo {author} {\bibfnamefont {J.}~\bibnamefont
  {Junquera}},\ }\bibfield  {title} {\bibinfo {title} {First-principles study
  of two-dimensional electron and hole gases at the head-to-head and
  tail-to-tail ${180}^{\ensuremath{\circ}}$ domain walls in
  {${\mathrm{PbTiO}}_{3}$} ferroelectric thin films},\ }\href
  {https://doi.org/10.1103/PhysRevB.101.174114} {\bibfield  {journal} {\bibinfo
   {journal} {Phys. Rev. B}\ }\textbf {\bibinfo {volume} {101}},\ \bibinfo
  {pages} {174114} (\bibinfo {year} {2020})}\BibitemShut {NoStop}%
\bibitem [{\citenamefont {Verhoff}\ \emph {et~al.}(2024)\citenamefont
  {Verhoff}, \citenamefont {Pionteck}, \citenamefont {R{\"u}sing},
  \citenamefont {Fritze}, \citenamefont {Eng},\ and\ \citenamefont
  {Sanna}}]{verhoff2024DFT}%
  \BibitemOpen
  \bibfield  {author} {\bibinfo {author} {\bibfnamefont {L.~M.}\ \bibnamefont
  {Verhoff}}, \bibinfo {author} {\bibfnamefont {M.~N.}\ \bibnamefont
  {Pionteck}}, \bibinfo {author} {\bibfnamefont {M.}~\bibnamefont
  {R{\"u}sing}}, \bibinfo {author} {\bibfnamefont {H.}~\bibnamefont {Fritze}},
  \bibinfo {author} {\bibfnamefont {L.~M.}\ \bibnamefont {Eng}},\ and\ \bibinfo
  {author} {\bibfnamefont {S.}~\bibnamefont {Sanna}},\ }\bibfield  {title}
  {\bibinfo {title} {Two-dimensional electronic conductivity in insulating
  ferroelectrics: Peculiar properties of domain walls},\ }\href@noop {}
  {\bibfield  {journal} {\bibinfo  {journal} {Physical Review Research}\
  }\textbf {\bibinfo {volume} {6}},\ \bibinfo {pages} {L042015} (\bibinfo
  {year} {2024})}\BibitemShut {NoStop}%
\bibitem [{\citenamefont {Lubk}\ \emph {et~al.}(2009)\citenamefont {Lubk},
  \citenamefont {Gemming},\ and\ \citenamefont {Spaldin}}]{Lubk:2009DFT}%
  \BibitemOpen
  \bibfield  {author} {\bibinfo {author} {\bibfnamefont {A.}~\bibnamefont
  {Lubk}}, \bibinfo {author} {\bibfnamefont {S.}~\bibnamefont {Gemming}},\ and\
  \bibinfo {author} {\bibfnamefont {N.~A.}\ \bibnamefont {Spaldin}},\
  }\bibfield  {title} {\bibinfo {title} {First-principles study of
  ferroelectric domain walls in multiferroic bismuth ferrite},\ }\href
  {https://doi.org/10.1103/PhysRevB.80.104110} {\bibfield  {journal} {\bibinfo
  {journal} {Phys. Rev. B}\ }\textbf {\bibinfo {volume} {80}},\ \bibinfo
  {pages} {104110} (\bibinfo {year} {2009})}\BibitemShut {NoStop}%
\bibitem [{\citenamefont {Zahn}\ \emph {et~al.}(2024)\citenamefont {Zahn},
  \citenamefont {Beyreuther}, \citenamefont {Kiseleva}, \citenamefont {Lotfy},
  \citenamefont {McCluskey}, \citenamefont {Maguire}, \citenamefont {Suna},
  \citenamefont {R\"using}, \citenamefont {Gregg},\ and\ \citenamefont
  {Eng}}]{Zahn:2024}%
  \BibitemOpen
  \bibfield  {author} {\bibinfo {author} {\bibfnamefont {M.}~\bibnamefont
  {Zahn}}, \bibinfo {author} {\bibfnamefont {E.}~\bibnamefont {Beyreuther}},
  \bibinfo {author} {\bibfnamefont {I.}~\bibnamefont {Kiseleva}}, \bibinfo
  {author} {\bibfnamefont {A.~S.}\ \bibnamefont {Lotfy}}, \bibinfo {author}
  {\bibfnamefont {C.~J.}\ \bibnamefont {McCluskey}}, \bibinfo {author}
  {\bibfnamefont {J.~R.}\ \bibnamefont {Maguire}}, \bibinfo {author}
  {\bibfnamefont {A.}~\bibnamefont {Suna}}, \bibinfo {author} {\bibfnamefont
  {M.}~\bibnamefont {R\"using}}, \bibinfo {author} {\bibfnamefont {J.~M.}\
  \bibnamefont {Gregg}},\ and\ \bibinfo {author} {\bibfnamefont {L.~M.}\
  \bibnamefont {Eng}},\ }\bibfield  {title} {\bibinfo {title}
  {Equivalent-circuit model that quantitatively describes domain-wall
  conductivity in ferroelectric {LiNbO$_3$}},\ }\href
  {https://doi.org/10.1103/PhysRevApplied.21.024007} {\bibfield  {journal}
  {\bibinfo  {journal} {Phys. Rev. Appl.}\ }\textbf {\bibinfo {volume} {21}},\
  \bibinfo {pages} {024007} (\bibinfo {year} {2024})}\BibitemShut {NoStop}%
\bibitem [{\citenamefont {Zhong}\ \emph {et~al.}(2013)\citenamefont {Zhong},
  \citenamefont {T{\'o}th},\ and\ \citenamefont {Held}}]{zhong2013theory}%
  \BibitemOpen
  \bibfield  {author} {\bibinfo {author} {\bibfnamefont {Z.}~\bibnamefont
  {Zhong}}, \bibinfo {author} {\bibfnamefont {A.}~\bibnamefont {T{\'o}th}},\
  and\ \bibinfo {author} {\bibfnamefont {K.}~\bibnamefont {Held}},\ }\bibfield
  {title} {\bibinfo {title} {Theory of spin-orbit coupling at
  {LaAlO$_3$/SrTiO$_3$} interfaces and {SrTiO$_3$} surfaces},\ }\href@noop {}
  {\bibfield  {journal} {\bibinfo  {journal} {Physical Review B}\ }\textbf
  {\bibinfo {volume} {87}},\ \bibinfo {pages} {161102} (\bibinfo {year}
  {2013})}\BibitemShut {NoStop}%
\bibitem [{\citenamefont {Tao}\ and\ \citenamefont
  {Wang}(2016)}]{tao2016strain}%
  \BibitemOpen
  \bibfield  {author} {\bibinfo {author} {\bibfnamefont {L.}~\bibnamefont
  {Tao}}\ and\ \bibinfo {author} {\bibfnamefont {J.}~\bibnamefont {Wang}},\
  }\bibfield  {title} {\bibinfo {title} {Strain-tunable ferroelectricity and
  its control of {Rashba} effect in {KTaO$_3$}},\ }\href@noop {} {\bibfield
  {journal} {\bibinfo  {journal} {Journal of Applied Physics}\ }\textbf
  {\bibinfo {volume} {120}} (\bibinfo {year} {2016})}\BibitemShut {NoStop}%
\bibitem [{\citenamefont {Bruno}\ \emph {et~al.}(2019)\citenamefont {Bruno},
  \citenamefont {McKeown~Walker}, \citenamefont {Ricc{\`o}}, \citenamefont
  {De~La~Torre}, \citenamefont {Wang}, \citenamefont {Tamai}, \citenamefont
  {Kim}, \citenamefont {Hoesch}, \citenamefont {Bahramy},\ and\ \citenamefont
  {Baumberger}}]{bruno2019band}%
  \BibitemOpen
  \bibfield  {author} {\bibinfo {author} {\bibfnamefont {F.~Y.}\ \bibnamefont
  {Bruno}}, \bibinfo {author} {\bibfnamefont {S.}~\bibnamefont
  {McKeown~Walker}}, \bibinfo {author} {\bibfnamefont {S.}~\bibnamefont
  {Ricc{\`o}}}, \bibinfo {author} {\bibfnamefont {A.}~\bibnamefont
  {De~La~Torre}}, \bibinfo {author} {\bibfnamefont {Z.}~\bibnamefont {Wang}},
  \bibinfo {author} {\bibfnamefont {A.}~\bibnamefont {Tamai}}, \bibinfo
  {author} {\bibfnamefont {T.~K.}\ \bibnamefont {Kim}}, \bibinfo {author}
  {\bibfnamefont {M.}~\bibnamefont {Hoesch}}, \bibinfo {author} {\bibfnamefont
  {M.~S.}\ \bibnamefont {Bahramy}},\ and\ \bibinfo {author} {\bibfnamefont
  {F.}~\bibnamefont {Baumberger}},\ }\bibfield  {title} {\bibinfo {title} {Band
  structure and spin--orbital texture of the (111)-{KTaO$_3$} {2D} electron
  gas},\ }\href@noop {} {\bibfield  {journal} {\bibinfo  {journal} {Advanced
  electronic materials}\ }\textbf {\bibinfo {volume} {5}},\ \bibinfo {pages}
  {1800860} (\bibinfo {year} {2019})}\BibitemShut {NoStop}%
\bibitem [{\citenamefont {Johansson}\ \emph {et~al.}(2021)\citenamefont
  {Johansson}, \citenamefont {G{\"o}bel}, \citenamefont {Henk}, \citenamefont
  {Bibes},\ and\ \citenamefont {Mertig}}]{johansson2021spin}%
  \BibitemOpen
  \bibfield  {author} {\bibinfo {author} {\bibfnamefont {A.}~\bibnamefont
  {Johansson}}, \bibinfo {author} {\bibfnamefont {B.}~\bibnamefont
  {G{\"o}bel}}, \bibinfo {author} {\bibfnamefont {J.}~\bibnamefont {Henk}},
  \bibinfo {author} {\bibfnamefont {M.}~\bibnamefont {Bibes}},\ and\ \bibinfo
  {author} {\bibfnamefont {I.}~\bibnamefont {Mertig}},\ }\bibfield  {title}
  {\bibinfo {title} {Spin and orbital {Edelstein} effects in a two-dimensional
  electron gas: Theory and application to srtio 3 interfaces},\ }\href@noop {}
  {\bibfield  {journal} {\bibinfo  {journal} {Physical Review Research}\
  }\textbf {\bibinfo {volume} {3}},\ \bibinfo {pages} {013275} (\bibinfo {year}
  {2021})}\BibitemShut {NoStop}%
\bibitem [{\citenamefont {Kumar}\ and\ \citenamefont
  {Ganguli}(2022)}]{kumar2022rashba}%
  \BibitemOpen
  \bibfield  {author} {\bibinfo {author} {\bibfnamefont {V.}~\bibnamefont
  {Kumar}}\ and\ \bibinfo {author} {\bibfnamefont {N.}~\bibnamefont
  {Ganguli}},\ }\bibfield  {title} {\bibinfo {title} {Rashba-like spin-orbit
  interaction and spin texture at the {KTaO$_3$} (001) surface from {DFT}
  calculations},\ }\href@noop {} {\bibfield  {journal} {\bibinfo  {journal}
  {Physical Review B}\ }\textbf {\bibinfo {volume} {106}},\ \bibinfo {pages}
  {125127} (\bibinfo {year} {2022})}\BibitemShut {NoStop}%
\bibitem [{\citenamefont {Trama}\ \emph {et~al.}(2022)\citenamefont {Trama},
  \citenamefont {Cataudella}, \citenamefont {Perroni}, \citenamefont {Romeo},\
  and\ \citenamefont {Citro}}]{trama2022tunable}%
  \BibitemOpen
  \bibfield  {author} {\bibinfo {author} {\bibfnamefont {M.}~\bibnamefont
  {Trama}}, \bibinfo {author} {\bibfnamefont {V.}~\bibnamefont {Cataudella}},
  \bibinfo {author} {\bibfnamefont {C.~A.}\ \bibnamefont {Perroni}}, \bibinfo
  {author} {\bibfnamefont {F.}~\bibnamefont {Romeo}},\ and\ \bibinfo {author}
  {\bibfnamefont {R.}~\bibnamefont {Citro}},\ }\bibfield  {title} {\bibinfo
  {title} {Tunable spin and orbital {Edelstein} effect at (111)
  {LaAlO$_3$/SrTiO$_3$} interface},\ }\href@noop {} {\bibfield  {journal}
  {\bibinfo  {journal} {Nanomaterials}\ }\textbf {\bibinfo {volume} {12}},\
  \bibinfo {pages} {2494} (\bibinfo {year} {2022})}\BibitemShut {NoStop}%
\bibitem [{\citenamefont {Han}\ \emph {et~al.}(2018)\citenamefont {Han},
  \citenamefont {Otani},\ and\ \citenamefont {Maekawa}}]{han2018quantum}%
  \BibitemOpen
  \bibfield  {author} {\bibinfo {author} {\bibfnamefont {W.}~\bibnamefont
  {Han}}, \bibinfo {author} {\bibfnamefont {Y.}~\bibnamefont {Otani}},\ and\
  \bibinfo {author} {\bibfnamefont {S.}~\bibnamefont {Maekawa}},\ }\bibfield
  {title} {\bibinfo {title} {Quantum materials for spin and charge
  conversion},\ }\href@noop {} {\bibfield  {journal} {\bibinfo  {journal} {npj
  Quantum Materials}\ }\textbf {\bibinfo {volume} {3}},\ \bibinfo {pages} {27}
  (\bibinfo {year} {2018})}\BibitemShut {NoStop}%
\bibitem [{\citenamefont {Vicente-Arche}\ \emph {et~al.}(2021)\citenamefont
  {Vicente-Arche}, \citenamefont {Br{\'e}hin}, \citenamefont {Varotto},
  \citenamefont {Cosset-Cheneau}, \citenamefont {Mallik}, \citenamefont
  {Salazar}, \citenamefont {No{\"e}l}, \citenamefont {Vaz}, \citenamefont
  {Trier}, \citenamefont {Bhattacharya} \emph {et~al.}}]{vicente2021spin}%
  \BibitemOpen
  \bibfield  {author} {\bibinfo {author} {\bibfnamefont {L.~M.}\ \bibnamefont
  {Vicente-Arche}}, \bibinfo {author} {\bibfnamefont {J.}~\bibnamefont
  {Br{\'e}hin}}, \bibinfo {author} {\bibfnamefont {S.}~\bibnamefont {Varotto}},
  \bibinfo {author} {\bibfnamefont {M.}~\bibnamefont {Cosset-Cheneau}},
  \bibinfo {author} {\bibfnamefont {S.}~\bibnamefont {Mallik}}, \bibinfo
  {author} {\bibfnamefont {R.}~\bibnamefont {Salazar}}, \bibinfo {author}
  {\bibfnamefont {P.}~\bibnamefont {No{\"e}l}}, \bibinfo {author}
  {\bibfnamefont {D.~C.}\ \bibnamefont {Vaz}}, \bibinfo {author} {\bibfnamefont
  {F.}~\bibnamefont {Trier}}, \bibinfo {author} {\bibfnamefont
  {S.}~\bibnamefont {Bhattacharya}}, \emph {et~al.},\ }\bibfield  {title}
  {\bibinfo {title} {Spin--charge interconversion in {KTaO$_3$} {2D} electron
  gases},\ }\href@noop {} {\bibfield  {journal} {\bibinfo  {journal} {Advanced
  Materials}\ }\textbf {\bibinfo {volume} {33}},\ \bibinfo {pages} {2102102}
  (\bibinfo {year} {2021})}\BibitemShut {NoStop}%
\bibitem [{\citenamefont {Al-Tawhid}\ \emph {et~al.}(2025)\citenamefont
  {Al-Tawhid}, \citenamefont {Sun}, \citenamefont {Comstock}, \citenamefont
  {Kumah}, \citenamefont {Sun},\ and\ \citenamefont {Ahadi}}]{al2025spin}%
  \BibitemOpen
  \bibfield  {author} {\bibinfo {author} {\bibfnamefont {A.~H.}\ \bibnamefont
  {Al-Tawhid}}, \bibinfo {author} {\bibfnamefont {R.}~\bibnamefont {Sun}},
  \bibinfo {author} {\bibfnamefont {A.~H.}\ \bibnamefont {Comstock}}, \bibinfo
  {author} {\bibfnamefont {D.~P.}\ \bibnamefont {Kumah}}, \bibinfo {author}
  {\bibfnamefont {D.}~\bibnamefont {Sun}},\ and\ \bibinfo {author}
  {\bibfnamefont {K.}~\bibnamefont {Ahadi}},\ }\bibfield  {title} {\bibinfo
  {title} {Spin-to-charge conversion at {KTaO$_3$} (111) interfaces},\
  }\href@noop {} {\bibfield  {journal} {\bibinfo  {journal} {Applied Physics
  Letters}\ }\textbf {\bibinfo {volume} {126}} (\bibinfo {year}
  {2025})}\BibitemShut {NoStop}%
\bibitem [{\citenamefont {Zhang}\ \emph
  {et~al.}(2019{\natexlab{b}})\citenamefont {Zhang}, \citenamefont {Ma},
  \citenamefont {Zhang}, \citenamefont {Chen}, \citenamefont {Wang},
  \citenamefont {Li}, \citenamefont {Yun}, \citenamefont {Yan}, \citenamefont
  {Chen}, \citenamefont {Hu} \emph {et~al.}}]{zhang2019thermal}%
  \BibitemOpen
  \bibfield  {author} {\bibinfo {author} {\bibfnamefont {H.}~\bibnamefont
  {Zhang}}, \bibinfo {author} {\bibfnamefont {Y.}~\bibnamefont {Ma}}, \bibinfo
  {author} {\bibfnamefont {H.}~\bibnamefont {Zhang}}, \bibinfo {author}
  {\bibfnamefont {X.}~\bibnamefont {Chen}}, \bibinfo {author} {\bibfnamefont
  {S.}~\bibnamefont {Wang}}, \bibinfo {author} {\bibfnamefont {G.}~\bibnamefont
  {Li}}, \bibinfo {author} {\bibfnamefont {Y.}~\bibnamefont {Yun}}, \bibinfo
  {author} {\bibfnamefont {X.}~\bibnamefont {Yan}}, \bibinfo {author}
  {\bibfnamefont {Y.}~\bibnamefont {Chen}}, \bibinfo {author} {\bibfnamefont
  {F.}~\bibnamefont {Hu}}, \emph {et~al.},\ }\bibfield  {title} {\bibinfo
  {title} {Thermal spin injection and inverse {Edelstein} effect of the
  two-dimensional electron gas at {EuO--KTaO$_3$} interfaces},\ }\href@noop {}
  {\bibfield  {journal} {\bibinfo  {journal} {Nano Letters}\ }\textbf {\bibinfo
  {volume} {19}},\ \bibinfo {pages} {1605} (\bibinfo {year}
  {2019}{\natexlab{b}})}\BibitemShut {NoStop}%
\bibitem [{\citenamefont {Lesne}\ \emph {et~al.}(2016)\citenamefont {Lesne},
  \citenamefont {Fu}, \citenamefont {Oyarz{\'u}n}, \citenamefont
  {Rojas-S{\'a}nchez}, \citenamefont {Vaz}, \citenamefont {Naganuma},
  \citenamefont {Sicoli}, \citenamefont {Attan{\'e}}, \citenamefont {Jamet},
  \citenamefont {Jacquet} \emph {et~al.}}]{lesne2016highly}%
  \BibitemOpen
  \bibfield  {author} {\bibinfo {author} {\bibfnamefont {E.}~\bibnamefont
  {Lesne}}, \bibinfo {author} {\bibfnamefont {Y.}~\bibnamefont {Fu}}, \bibinfo
  {author} {\bibfnamefont {S.}~\bibnamefont {Oyarz{\'u}n}}, \bibinfo {author}
  {\bibfnamefont {J.~C.}\ \bibnamefont {Rojas-S{\'a}nchez}}, \bibinfo {author}
  {\bibfnamefont {D.}~\bibnamefont {Vaz}}, \bibinfo {author} {\bibfnamefont
  {H.}~\bibnamefont {Naganuma}}, \bibinfo {author} {\bibfnamefont
  {G.}~\bibnamefont {Sicoli}}, \bibinfo {author} {\bibfnamefont {J.-P.}\
  \bibnamefont {Attan{\'e}}}, \bibinfo {author} {\bibfnamefont
  {M.}~\bibnamefont {Jamet}}, \bibinfo {author} {\bibfnamefont
  {E.}~\bibnamefont {Jacquet}}, \emph {et~al.},\ }\bibfield  {title} {\bibinfo
  {title} {Highly efficient and tunable spin-to-charge conversion through
  {Rashba} coupling at oxide interfaces},\ }\href@noop {} {\bibfield  {journal}
  {\bibinfo  {journal} {Nature materials}\ }\textbf {\bibinfo {volume} {15}},\
  \bibinfo {pages} {1261} (\bibinfo {year} {2016})}\BibitemShut {NoStop}%
\bibitem [{\citenamefont {Song}\ \emph {et~al.}(2017)\citenamefont {Song},
  \citenamefont {Zhang}, \citenamefont {Su}, \citenamefont {Yuan},
  \citenamefont {Chen}, \citenamefont {Xing}, \citenamefont {Shi},
  \citenamefont {Sun},\ and\ \citenamefont {Han}}]{song2017observation}%
  \BibitemOpen
  \bibfield  {author} {\bibinfo {author} {\bibfnamefont {Q.}~\bibnamefont
  {Song}}, \bibinfo {author} {\bibfnamefont {H.}~\bibnamefont {Zhang}},
  \bibinfo {author} {\bibfnamefont {T.}~\bibnamefont {Su}}, \bibinfo {author}
  {\bibfnamefont {W.}~\bibnamefont {Yuan}}, \bibinfo {author} {\bibfnamefont
  {Y.}~\bibnamefont {Chen}}, \bibinfo {author} {\bibfnamefont {W.}~\bibnamefont
  {Xing}}, \bibinfo {author} {\bibfnamefont {J.}~\bibnamefont {Shi}}, \bibinfo
  {author} {\bibfnamefont {J.}~\bibnamefont {Sun}},\ and\ \bibinfo {author}
  {\bibfnamefont {W.}~\bibnamefont {Han}},\ }\bibfield  {title} {\bibinfo
  {title} {Observation of inverse {Edelstein} effect in {Rashba}-split {2DEG}
  between {SrTiO$_3$} and {LaAlO$_3$} at room temperature},\ }\href@noop {}
  {\bibfield  {journal} {\bibinfo  {journal} {Science advances}\ }\textbf
  {\bibinfo {volume} {3}},\ \bibinfo {pages} {e1602312} (\bibinfo {year}
  {2017})}\BibitemShut {NoStop}%
\bibitem [{\citenamefont {No{\"e}l}\ \emph {et~al.}(2020)\citenamefont
  {No{\"e}l}, \citenamefont {Trier}, \citenamefont {Vicente~Arche},
  \citenamefont {Br{\'e}hin}, \citenamefont {Vaz}, \citenamefont {Garcia},
  \citenamefont {Fusil}, \citenamefont {Barth{\'e}l{\'e}my}, \citenamefont
  {Vila}, \citenamefont {Bibes} \emph {et~al.}}]{noel2020non}%
  \BibitemOpen
  \bibfield  {author} {\bibinfo {author} {\bibfnamefont {P.}~\bibnamefont
  {No{\"e}l}}, \bibinfo {author} {\bibfnamefont {F.}~\bibnamefont {Trier}},
  \bibinfo {author} {\bibfnamefont {L.~M.}\ \bibnamefont {Vicente~Arche}},
  \bibinfo {author} {\bibfnamefont {J.}~\bibnamefont {Br{\'e}hin}}, \bibinfo
  {author} {\bibfnamefont {D.~C.}\ \bibnamefont {Vaz}}, \bibinfo {author}
  {\bibfnamefont {V.}~\bibnamefont {Garcia}}, \bibinfo {author} {\bibfnamefont
  {S.}~\bibnamefont {Fusil}}, \bibinfo {author} {\bibfnamefont
  {A.}~\bibnamefont {Barth{\'e}l{\'e}my}}, \bibinfo {author} {\bibfnamefont
  {L.}~\bibnamefont {Vila}}, \bibinfo {author} {\bibfnamefont {M.}~\bibnamefont
  {Bibes}}, \emph {et~al.},\ }\bibfield  {title} {\bibinfo {title}
  {Non-volatile electric control of spin--charge conversion in a {SrTiO$_3$}
  {Rashba} system},\ }\href@noop {} {\bibfield  {journal} {\bibinfo  {journal}
  {Nature}\ }\textbf {\bibinfo {volume} {580}},\ \bibinfo {pages} {483}
  (\bibinfo {year} {2020})}\BibitemShut {NoStop}%
\bibitem [{\citenamefont {El~Hamdi}\ \emph {et~al.}(2023)\citenamefont
  {El~Hamdi}, \citenamefont {Chauleau}, \citenamefont {Boselli}, \citenamefont
  {Thibault}, \citenamefont {Gorini}, \citenamefont {Smogunov}, \citenamefont
  {Barreteau}, \citenamefont {Gariglio}, \citenamefont {Triscone},\ and\
  \citenamefont {Viret}}]{el2023observation}%
  \BibitemOpen
  \bibfield  {author} {\bibinfo {author} {\bibfnamefont {A.}~\bibnamefont
  {El~Hamdi}}, \bibinfo {author} {\bibfnamefont {J.-Y.}\ \bibnamefont
  {Chauleau}}, \bibinfo {author} {\bibfnamefont {M.}~\bibnamefont {Boselli}},
  \bibinfo {author} {\bibfnamefont {C.}~\bibnamefont {Thibault}}, \bibinfo
  {author} {\bibfnamefont {C.}~\bibnamefont {Gorini}}, \bibinfo {author}
  {\bibfnamefont {A.}~\bibnamefont {Smogunov}}, \bibinfo {author}
  {\bibfnamefont {C.}~\bibnamefont {Barreteau}}, \bibinfo {author}
  {\bibfnamefont {S.}~\bibnamefont {Gariglio}}, \bibinfo {author}
  {\bibfnamefont {J.-M.}\ \bibnamefont {Triscone}},\ and\ \bibinfo {author}
  {\bibfnamefont {M.}~\bibnamefont {Viret}},\ }\bibfield  {title} {\bibinfo
  {title} {Observation of the orbital inverse {Rashba--Edelstein} effect},\
  }\href@noop {} {\bibfield  {journal} {\bibinfo  {journal} {Nature Physics}\
  }\textbf {\bibinfo {volume} {19}},\ \bibinfo {pages} {1855} (\bibinfo {year}
  {2023})}\BibitemShut {NoStop}%
\bibitem [{\citenamefont {Gureev}\ \emph {et~al.}(2011)\citenamefont {Gureev},
  \citenamefont {Tagantsev},\ and\ \citenamefont {Setter}}]{Gureev:2011}%
  \BibitemOpen
  \bibfield  {author} {\bibinfo {author} {\bibfnamefont {M.~Y.}\ \bibnamefont
  {Gureev}}, \bibinfo {author} {\bibfnamefont {A.~K.}\ \bibnamefont
  {Tagantsev}},\ and\ \bibinfo {author} {\bibfnamefont {N.}~\bibnamefont
  {Setter}},\ }\bibfield  {title} {\bibinfo {title} {Head-to-head and
  tail-to-tail 180$^\circ$ domain walls in an isolated ferroelectric},\
  }\href@noop {} {\bibfield  {journal} {\bibinfo  {journal} {Phys. Rev. B}\
  }\textbf {\bibinfo {volume} {83}},\ \bibinfo {pages} {184104} (\bibinfo
  {year} {2011})}\BibitemShut {NoStop}%
\bibitem [{\citenamefont {Sturman}\ \emph {et~al.}(2015)\citenamefont
  {Sturman}, \citenamefont {Podivilov}, \citenamefont {Stepanov}, \citenamefont
  {Tagantsev},\ and\ \citenamefont {Setter}}]{sturman_quantum_2015}%
  \BibitemOpen
  \bibfield  {author} {\bibinfo {author} {\bibfnamefont {B.}~\bibnamefont
  {Sturman}}, \bibinfo {author} {\bibfnamefont {E.}~\bibnamefont {Podivilov}},
  \bibinfo {author} {\bibfnamefont {M.}~\bibnamefont {Stepanov}}, \bibinfo
  {author} {\bibfnamefont {A.}~\bibnamefont {Tagantsev}},\ and\ \bibinfo
  {author} {\bibfnamefont {N.}~\bibnamefont {Setter}},\ }\bibfield  {title}
  {\bibinfo {title} {Quantum properties of charged ferroelectric domain
  walls},\ }\href {https://doi.org/10.1103/PhysRevB.92.214112} {\bibfield
  {journal} {\bibinfo  {journal} {Phys. Rev. B}\ }\textbf {\bibinfo {volume}
  {92}},\ \bibinfo {pages} {214112} (\bibinfo {year} {2015})}\BibitemShut
  {NoStop}%
\bibitem [{\citenamefont {Chapman}\ and\ \citenamefont
  {Atkinson}(2022)}]{Chapman:2022}%
  \BibitemOpen
  \bibfield  {author} {\bibinfo {author} {\bibfnamefont {K.~S.}\ \bibnamefont
  {Chapman}}\ and\ \bibinfo {author} {\bibfnamefont {W.~A.}\ \bibnamefont
  {Atkinson}},\ }\bibfield  {title} {\bibinfo {title} {Mechanism for
  switchability in electron-doped ferroelectric interfaces},\ }\href
  {https://doi.org/10.1103/PhysRevB.105.035307} {\bibfield  {journal} {\bibinfo
   {journal} {Phys. Rev. B}\ }\textbf {\bibinfo {volume} {105}},\ \bibinfo
  {pages} {035307} (\bibinfo {year} {2022})}\BibitemShut {NoStop}%
\bibitem [{\citenamefont {Atkinson}(2022)}]{atkinson2022evolution}%
  \BibitemOpen
  \bibfield  {author} {\bibinfo {author} {\bibfnamefont {W.}~\bibnamefont
  {Atkinson}},\ }\bibfield  {title} {\bibinfo {title} {Evolution of domain
  structure with electron doping in ferroelectric thin films},\ }\href@noop {}
  {\bibfield  {journal} {\bibinfo  {journal} {Phys. Rev. B}\ }\textbf {\bibinfo
  {volume} {106}},\ \bibinfo {pages} {134102} (\bibinfo {year}
  {2022})}\BibitemShut {NoStop}%
\bibitem [{\citenamefont {Cornell}\ and\ \citenamefont
  {Atkinson}(2023)}]{cornell2023influence}%
  \BibitemOpen
  \bibfield  {author} {\bibinfo {author} {\bibfnamefont {B.~C.}\ \bibnamefont
  {Cornell}}\ and\ \bibinfo {author} {\bibfnamefont {W.~A.}\ \bibnamefont
  {Atkinson}},\ }\bibfield  {title} {\bibinfo {title} {Influence of a realistic
  multiorbital band structure on conducting domain walls in perovskite
  ferroelectrics},\ }\href@noop {} {\bibfield  {journal} {\bibinfo  {journal}
  {Phys. Rev. B}\ }\textbf {\bibinfo {volume} {108}},\ \bibinfo {pages}
  {245108} (\bibinfo {year} {2023})}\BibitemShut {NoStop}%
\bibitem [{\citenamefont {Salehi}\ \emph {et~al.}(2003)\citenamefont {Salehi},
  \citenamefont {Shahtahmasebi},\ and\ \citenamefont
  {Hosseini}}]{salehi2003band}%
  \BibitemOpen
  \bibfield  {author} {\bibinfo {author} {\bibfnamefont {H.}~\bibnamefont
  {Salehi}}, \bibinfo {author} {\bibfnamefont {N.}~\bibnamefont
  {Shahtahmasebi}},\ and\ \bibinfo {author} {\bibfnamefont {S.}~\bibnamefont
  {Hosseini}},\ }\bibfield  {title} {\bibinfo {title} {Band structure of
  tetragonal {BaTiO$_3$}},\ }\href@noop {} {\bibfield  {journal} {\bibinfo
  {journal} {The European Physical Journal B-Condensed Matter and Complex
  Systems}\ }\textbf {\bibinfo {volume} {32}},\ \bibinfo {pages} {177}
  (\bibinfo {year} {2003})}\BibitemShut {NoStop}%
\bibitem [{\citenamefont {Dunn}(1961)}]{TF9615701441}%
  \BibitemOpen
  \bibfield  {author} {\bibinfo {author} {\bibfnamefont {T.~M.}\ \bibnamefont
  {Dunn}},\ }\bibfield  {title} {\bibinfo {title} {Spin-orbit coupling in the
  first and second transition series},\ }\href
  {https://doi.org/10.1039/TF9615701441} {\bibfield  {journal} {\bibinfo
  {journal} {Trans. Faraday Soc.}\ }\textbf {\bibinfo {volume} {57}},\ \bibinfo
  {pages} {1441} (\bibinfo {year} {1961})}\BibitemShut {NoStop}%
\bibitem [{\citenamefont {Ishikawa}\ \emph {et~al.}(2019)\citenamefont
  {Ishikawa}, \citenamefont {Takayama}, \citenamefont {Kremer}, \citenamefont
  {Nuss}, \citenamefont {Dinnebier}, \citenamefont {Kitagawa}, \citenamefont
  {Ishii},\ and\ \citenamefont {Takagi}}]{Ishikawa:2019ASOC}%
  \BibitemOpen
  \bibfield  {author} {\bibinfo {author} {\bibfnamefont {H.}~\bibnamefont
  {Ishikawa}}, \bibinfo {author} {\bibfnamefont {T.}~\bibnamefont {Takayama}},
  \bibinfo {author} {\bibfnamefont {R.~K.}\ \bibnamefont {Kremer}}, \bibinfo
  {author} {\bibfnamefont {J.}~\bibnamefont {Nuss}}, \bibinfo {author}
  {\bibfnamefont {R.}~\bibnamefont {Dinnebier}}, \bibinfo {author}
  {\bibfnamefont {K.}~\bibnamefont {Kitagawa}}, \bibinfo {author}
  {\bibfnamefont {K.}~\bibnamefont {Ishii}},\ and\ \bibinfo {author}
  {\bibfnamefont {H.}~\bibnamefont {Takagi}},\ }\bibfield  {title} {\bibinfo
  {title} {Ordering of hidden multipoles in spin-orbit entangled $5{d}^{1}$ ta
  chlorides},\ }\href {https://doi.org/10.1103/PhysRevB.100.045142} {\bibfield
  {journal} {\bibinfo  {journal} {Phys. Rev. B}\ }\textbf {\bibinfo {volume}
  {100}},\ \bibinfo {pages} {045142} (\bibinfo {year} {2019})}\BibitemShut
  {NoStop}%
\bibitem [{\citenamefont {Vivek}\ \emph {et~al.}(2017)\citenamefont {Vivek},
  \citenamefont {Goerbig},\ and\ \citenamefont {Gabay}}]{vivek2017topological}%
  \BibitemOpen
  \bibfield  {author} {\bibinfo {author} {\bibfnamefont {M.}~\bibnamefont
  {Vivek}}, \bibinfo {author} {\bibfnamefont {M.~O.}\ \bibnamefont {Goerbig}},\
  and\ \bibinfo {author} {\bibfnamefont {M.}~\bibnamefont {Gabay}},\ }\bibfield
   {title} {\bibinfo {title} {Topological states at the (001) surface of srtio
  3},\ }\href@noop {} {\bibfield  {journal} {\bibinfo  {journal} {Physical
  Review B}\ }\textbf {\bibinfo {volume} {95}},\ \bibinfo {pages} {165117}
  (\bibinfo {year} {2017})}\BibitemShut {NoStop}%
\bibitem [{\citenamefont {Raslan}\ \emph {et~al.}(2017)\citenamefont {Raslan},
  \citenamefont {Lafleur},\ and\ \citenamefont
  {Atkinson}}]{Raslan:2017temperature}%
  \BibitemOpen
  \bibfield  {author} {\bibinfo {author} {\bibfnamefont {A.}~\bibnamefont
  {Raslan}}, \bibinfo {author} {\bibfnamefont {P.}~\bibnamefont {Lafleur}},\
  and\ \bibinfo {author} {\bibfnamefont {W.~A.}\ \bibnamefont {Atkinson}},\
  }\bibfield  {title} {\bibinfo {title} {Temperature-dependent band structure
  of {${\mathrm{SrTiO}}_{3}$} interfaces},\ }\href
  {https://doi.org/10.1103/PhysRevB.95.054106} {\bibfield  {journal} {\bibinfo
  {journal} {Phys. Rev. B}\ }\textbf {\bibinfo {volume} {95}},\ \bibinfo
  {pages} {054106} (\bibinfo {year} {2017})}\BibitemShut {NoStop}%
\bibitem [{\citenamefont {Hammar}\ and\ \citenamefont
  {Johnson}(2000)}]{hammar2000potentiometric}%
  \BibitemOpen
  \bibfield  {author} {\bibinfo {author} {\bibfnamefont {P.~R.}\ \bibnamefont
  {Hammar}}\ and\ \bibinfo {author} {\bibfnamefont {M.}~\bibnamefont
  {Johnson}},\ }\bibfield  {title} {\bibinfo {title} {Potentiometric
  measurements of the spin-split subbands in a two-dimensional electron gas},\
  }\href@noop {} {\bibfield  {journal} {\bibinfo  {journal} {Physical Review
  B}\ }\textbf {\bibinfo {volume} {61}},\ \bibinfo {pages} {7207} (\bibinfo
  {year} {2000})}\BibitemShut {NoStop}%
\end{thebibliography}

%

\end{document}